\documentclass[12pt,preprint,referee]{aastex}

\RequirePackage{lineno}
   \usepackage{color}
\usepackage{txfonts}
\usepackage{gensymb}

\newcommand{\be} {\begin{equation}}

\newcommand{\bc}{\begin{center}}
\newcommand{\ec}{\end{center}}
\def\ltsima{$\; \buildrel < \over \sim \;$}
\def\lsim{\lower.5ex\hbox{\ltsima}}
\def\loe{\lower.5ex\hbox{\ltsima}}
\def\gtsima{$\; \buildrel > \over \sim \;$}
\def\gsim{\lower.5ex\hbox{\gtsima}}
\def\goe{\lower.5ex\hbox{\gtsima}}

\def\ltsima{$\; \buildrel < \over \sim \;$}
\def\lsim{\lower.5ex\hbox{\ltsima}}
\def\loe{\lower.5ex\hbox{\ltsima}}
\def\gtsima{$\; \buildrel > \over \sim \;$}
\def\gsim{\lower.5ex\hbox{\gtsima}}
\def\goe{\lower.5ex\hbox{\gtsima}}

\def\ergscm2 {erg\,s$^{-1}$cm$^{-2}$}

\def\cm2 {cm$^{-2}$}

\usepackage{color}

\shortauthors{J. Li, D. F. Torres et al.}
\shorttitle{Non-thermal X-ray pulsars}

\begin{document}
\title{Theoretically motivated search and detection of non-thermal pulsations from {PSRs} J1747-2958, J2021+3651, and J1826-1256}

\author{Jian Li\altaffilmark{1}, Diego F. Torres\altaffilmark{2,3,4}, Francesco Coti Zelati\altaffilmark{2,3}, \\ Alessandro Papitto\altaffilmark{5}, Matthew Kerr\altaffilmark{6}, Nanda Rea\altaffilmark{2,3}}

\altaffiltext{1}{Deutsches Elektronen Synchrotron DESY, D-15738 Zeuthen, Germany;jian.li@desy.de}
\altaffiltext{2}{Institute of Space Sciences (ICE, CSIC), Campus UAB, Carrer de Can Magrans, 08193, Barcelona, Spain}
\altaffiltext{3}{Institut d'Estudis Espacials de Catalunya (IEEC), 08034 Barcelona, Spain}
\altaffiltext{4}{Instituci\'o Catalana de Recerca i Estudis Avan\c{c}ats (ICREA), E-08010, Barcelona, Spain; Email: dtorres@ice.csic.es}
\altaffiltext{5}{INAF-Osservatorio Astronomico di Roma, via di Frascati 33, I-00040 Monte Porzio Catone, Roma, Italy}
\altaffiltext{6}{Space Science Division, Naval Research Laboratory, Washington, DC 20375, USA}

\begin{abstract}

Based on a theoretical selection of pulsars as candidates for detection  at X-ray energies,
we present an analysis of archival X-ray observations done with {\it Chandra} and {\it XMM-Newton} of
PSR J1747-2958 (the pulsar in the `Mouse' nebula), PSR J2021+3651 (the pulsar in the `Dragonfly'
nebula), and PSR J1826-1256.
{X-ray pulsations from PSR J1747-2958 and PSR J1826-1256 are detected for the first time, and a previously reported hint of an X-ray pulsation from PSR J2021+3651 is confirmed {with a higher significance}.}
We analyze these pulsars' spectra in regards to the theoretically predicted energy distribution, finding a remarkable agreement, and provide here a refined calculation of the model parameters taking into account the newly derived X-ray spectral data.

\end{abstract}


\section{Introduction}
\label{intro}

In a recent article, we have introduced a model for the high-energy emission of pulsars  (Torres 2018).
According to this model, pulsed spectra detected in gamma-rays and/or X-rays is produced via synchro-curvature radiation, and can be described using just three physical parameters ({accelerating electric field, contrast and magnetic gradient}).
Interestingly, it was shown that if the model parameters were adjusted to describe the gamma-ray data, the resulting spectral energy distribution at lower X-ray energies is a relatively good representation of the spectra therein.
In particular, in all cases analyzed by Torres (2018) for which both X-ray and gamma-ray data were available, a fit only to the gamma-ray part of the spectra would miss the measured X-ray flux level by less than a factor of $\sim 2$.
This was the case even when one of the parameters --on which we comment below-- was fixed to an average value (gamma-ray data only is not enough constraining to fix it), and thus the spectral shape is completely determined by 2 {parameters}.
Thus, the model was posed to be a tool to distinguish which of the pulsars detected in gamma-rays could be {good candidates} to appear in X-rays. A list of pulsars appearing in the
Second \emph{Fermi}-LAT Pulsar Catalog (Abdo et al. 2013, 2PC hereafter)
for which the theoretically expected X-ray flux is  larger than 10$^{-13}$ erg cm$^{-2}$ s$^{-1}$ at 10 keV
was provided in Torres (2018).

Enlarging the non-thermal X-ray pulsar population is clearly an important task: it suffices to say that out of more than 200 gamma-ray pulsars known in the Galaxy, only {less than} 10\% have been detected in X-rays (see e.g., Kuiper and Hermsen 2015).
Enlarging the sample of X-ray pulsars is crucial to understand their global properties, and characterize how these compare with those of other pulsar subsamples.

In this Letter, we take three pulsars out of the list in Torres (2018) and analyze existing, {publicly available X-ray data} for them. We find that pulsed radiation is indeed found for all these pulsars, and provide an analysis of their spectral properties, comparing with model predictions taking into account (and not) the now determined X-ray data.

\section{Observation and data analysis}

We analyze archival {\it XMM-Newton} and {\it Chandra} data.
The {\it Chandra} data were reduced using CIAO version 4.7 and CALDB version 4.7.7.
We reprocessed the {\it Chandra} data to level=2 and removed periods of high background or flaring appearing in the observations.
{\it XMM-Newton} data were reduced with the {\it XMM-Newton} Science Analysis System (SAS, version 16.1.0).
Standard pipeline tasks (epproc for PN and emproc for MOS data) were used to process the raw observation data files (ODFs).
{\it XMM-Newton} data were also filtered to avoid the periods of hard X-ray background flares.
All the X-ray spectra were rebinned to have at least 25 counts for each channel.
%, as requested by the $\chi^{2}$ statistics.
%
Spectral analysis was carried out with XSPEC version 12.8.2.

From the spectral {modelling} of each pulsar we constructed the corresponding phase-averaged spectral energy distribution (SED).
These SEDs are produced by \emph{Xspec} using the corresponding unabsorbed spectral models (i.e., the corresponding fitted pulsar's power-law models in Table 1) and the real exposure times and background.

For timing analysis, all {time of arrivals of} X-ray photons were barycenter-corrected using the position of the pulsars and the latest JPL DE405 Earth ephemeris.
The pulsations were searched via an Z$^{2}_{n}$-test procedure around the corresponding periods {expected from the {\it Fermi}-LAT gamma-ray ephemeris or from previous publications},  with Fourier resolution (Buccheri et al. 1983), and with the number of harmonics $n$ varied from 1 to 5.
{We start from Z$^{2}_{1}$-test and {increase $n$ until a significant} signal is detected.
}%

{As we shall see below, the face-value periods we find in X-rays --considering the uncertainty coming only from the Fourier resolution--
are close, but not exactly at the gamma-ray pulsation period found earlier with {\it Fermi}-LAT.
In addition {to} intrinsic phenomena in the pulsars, and of timing noise being very large in at least two of the sources of our sample,
we may consider other possible uncertainties that may lead to such differences.}
{Among them, we consider the particular
instrument timing resolution and accuracy, possible fluctuations of observed counts, and of course also the Fourier resolution.}

To consider an uncertainty according to the corresponding timing resolution and accuracy,
the arrival time of each event was uniformly sampled within its corresponding uncertainty to produce a simulated event list, which was later analyzed via the Z$^{2}_{n}$-test.
This procedure was repeated 10000 times for each source{, leading to the estimation of a related period uncertainty.
Similarly, to estimate the period uncertainty resulting from fluctuations of observed counts, we sampled within the phase bins of the corresponding folded lightcurves, assuming a Poisson distribution.
Then the phases of sampled events are converted into arrival times in the observation to produce a simulated event list, which was again, later analyzed via the Z$^{2}_{n}$-test.
This process was also repeated 10000 times for each source, leading to the estimation of a related period uncertainty.}
The errors in the periods were estimated as the sum in quadrature of the {simulation obtained uncertainties} and the Fourier resolution,
{with the former dominating the period error.}
In the folded pulse profile, an arbitrary {reference epoch} $T_{0}$ was set to the start of respective observations.
{The pulse profile was modelled by a sinusoidal function with n harmonics, while n was decided by the Z$^{2}_{n}$-test.
The pulse fraction is defined as the ratio between the {semi-amplitude} of fundamental sinusoid and the average count rate.}
All the uncertainties {quoted} in the {spectral} analysis are at 1$\sigma$ confidence level.
Considering these uncertainties we find that the {\it Fermi}-LAT periods found earlier and the X-ray periods we detect are compatible in all three cases studied.

\subsection{PSR J1747-2958: Detection of the pulsar in the `Mouse' nebula}

J1747-2958 is a {98.8} ms young pulsar, detected in radio and gamma-rays (see, e.g., Camilo et al. 2002; 2PC) associated with the axisymmetric nebula G359.23-0.82 (referred to as
the Mouse).
The Mouse nebula has been discovered and studied in detail with {\it Chandra} (Gaensler et al. 2004; {Klingler et al. 2018}).
In {Gaensler et al. (2004)}, the head region of the Mouse nebula was decomposed into two Gaussian components. The first Gaussian was identified as
a point-like source possibly associated with the pulsar itself, albeit no X-ray pulsations were earlier reported.

{\it Chandra}/HRC observed PSR J1747-2958 with 58 ks exposure on Feb. 7th, 2008 (obs. ID 9106), providing sufficient timing {accuracy}({$\sim$ 11.1 ms, which is the median of time differences between observed events})\footnote{\url{http://cxc.harvard.edu/cal/Hrc/timing$\_$200304.html}}.
We considered this data set
and searched for pulsations  via the Z$^{2}_{n}$-test procedure.
Photons were extracted with a radius of 1 arcsec using the position of the first Gaussian reported in Gaensler et al. (2004) {which was proposed to be the pulsar magnetosphereic emission (Gaensler et al. 2004)}.
We found a peak at{ $P=0.09878(10)$ {s} (90\% uncertainty)} with a Z$^{2}_{2}$ statistic of 39.33 (Figure \ref{Z2}), which corresponds to a significance $\sim 5.3 \sigma$ after trials correction.
{The trials are the different values of n (1 and 2) considered in Z$^{2}_{n}$-test.
}%
The latest {\it Fermi}-LAT  gamma-ray ephemeris\footnote{LAT Gamma-ray Pulsar Timing Models,\url{https://confluence.slac.stanford.edu/display/GLAMCOG/LAT+Gamma-ray+Pulsar+Timing+Models}} covers from Aug. 13th, 2008 to Oct. 7th, 2013.
We extrapolated the gamma-ray ephemeris to the epoch of the X-ray observation.
The expected pulse period {is {$P=0.098823858(4)$ s}, which is compatible
within the uncertainty of the period we detected in X-rays.
{Additionally, we note that the timing noise is not considered in the pulse period prediction from gamma-ray ephemeris.
Based on current LAT Gamma-ray Pulsar Timing Models, the timing noise of PSR J1747-2958 may reach values as large as $\sim$ 2$\times$10$^{-5}$s during the X-ray observation, leading to additional uncertainties in the predicted period.}
The HRC data were folded at the detected period and the pulse profile is also shown in Figure \ref{Z2}, yielding a pulse fraction of {20.9\%$\pm$4.7\%.}
{Klingler et al. (2018) carried out a search of X-ray pulsations in the same {\it Chandra}/HRC data, but reported  no detection.
This is possibly due to different extraction regions, and/or a different number of events considered in the timing analysis.
However, we note that the pulse fraction derived in our analysis is consistent with the upper limit calculated by Klingler et al. (2018) (34\%) .}

We carried out the spectral analysis of PSR J1747-2958 using archival {\it Chandra} observations (ID 14519, 14520, 14521, 14522).
The pulsar spectrum was extracted from an elliptical region centered on the first Gaussian component reported in Gaensler et al. (2004), having a
major axis of 1.1 arcsec, ellipticity of 0.25, and a position angle of 118$\degree$ (see Figure \ref{maps}, top left panel).
To exclude the contamination of the nebula, the background spectrum was extracted from an elliptical region centered on the second Gaussian component with major axis of 2.4 arcsec, ellipticity of 0.12 and a position angle of 91$\degree$, {excluding}
the region used for the pulsar spectrum extraction.
{To increase the statistics, spectra from different observations are combined using the task \emph{combine\_spectra}}.
{We have tested that a simultaneous fitting to the spectra from different observations leads to consistent results}.
The combined spectra of PSR J1747-2958 could be well fitted with an absorbed power-law (see Figure \ref{spec}, reduced $\chi^{2}$=1.16, {D.O.F=56}).
The fitted N$_{H}$ and flux level (Table 1) are consistent with the value reported in Gaensler et al. (2004).
An absorbed blackbody could not lead to an acceptable fit (reduced $\chi^{2}$=1.7, {D.O.F=56}).
We also tested {an} absorbed power-law plus blackbody {model}.
However, the blackbody component is not significantly required according to the F-test (significance below 3$\sigma$).
Thus, we conclude that the spectrum we observe from PSR J1747-2958 is non-thermal.
%

%%%%%%%%%%%%%%%%%%%%%%%%%%%%%%%%%%%
\begin{center}
\begin{figure*}
\centering
\includegraphics[scale=0.4]{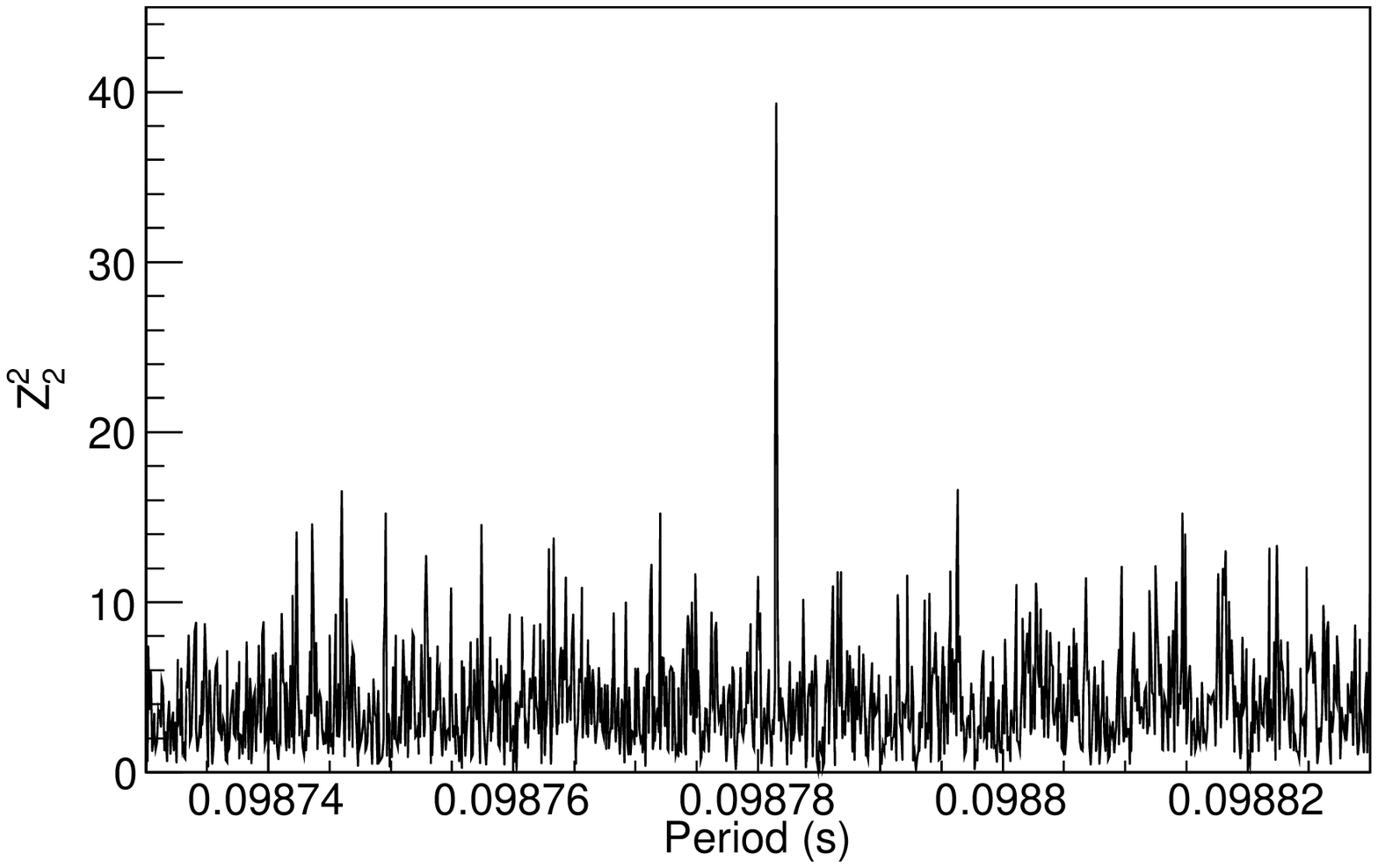}
\includegraphics[scale=0.4]{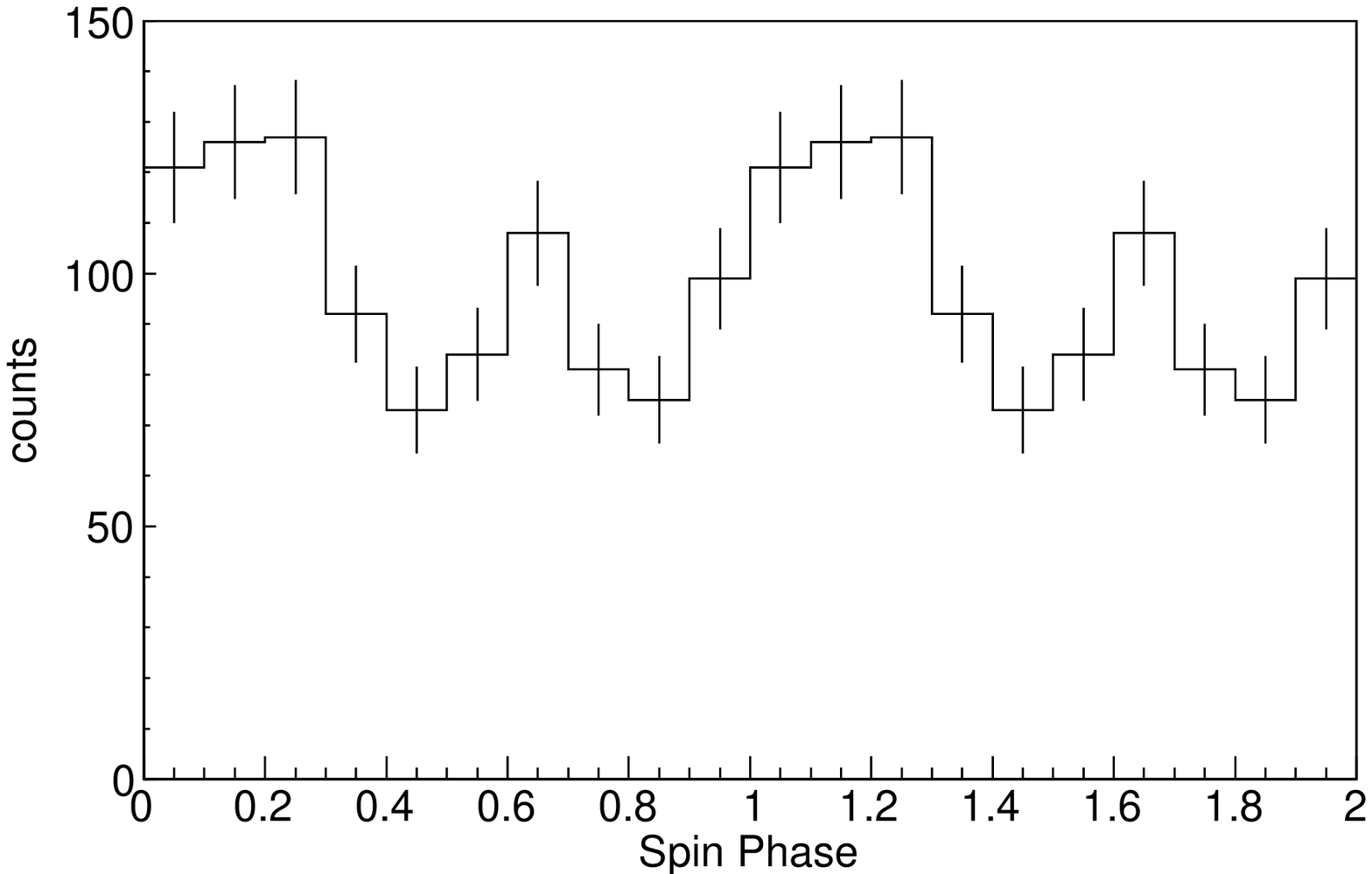}
\includegraphics[scale=0.4]{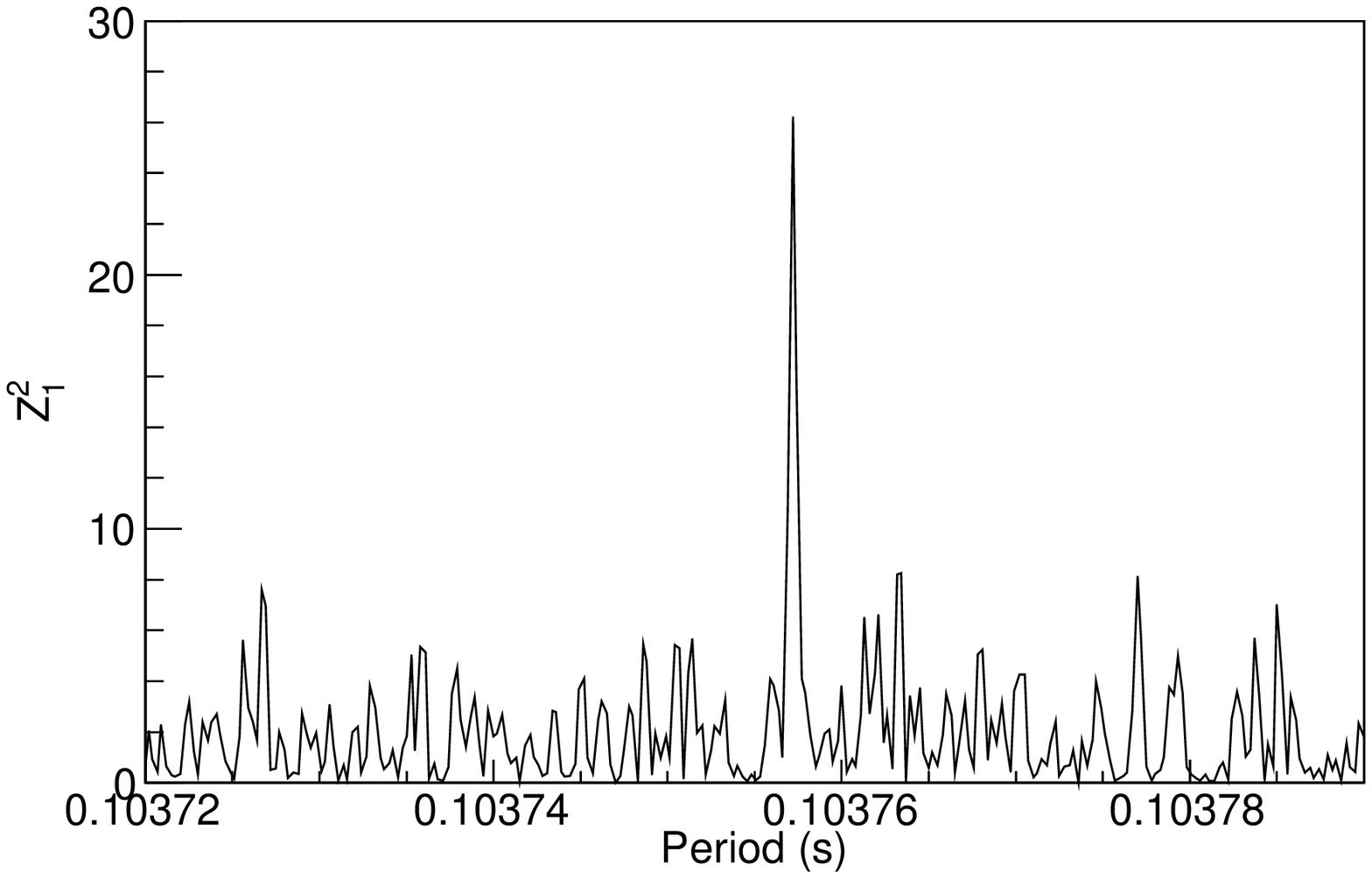}
\includegraphics[scale=0.4]{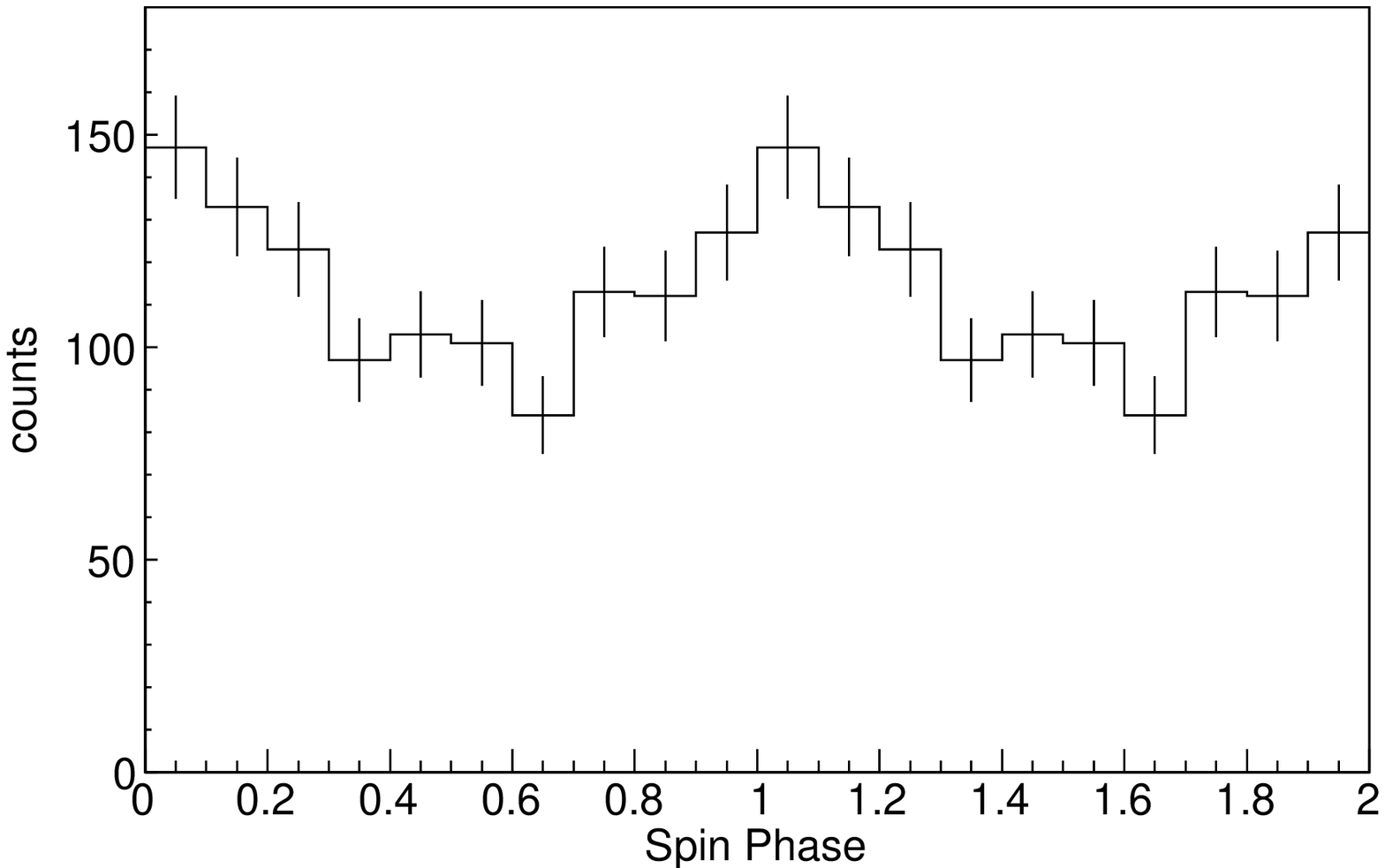}
\includegraphics[scale=0.4]{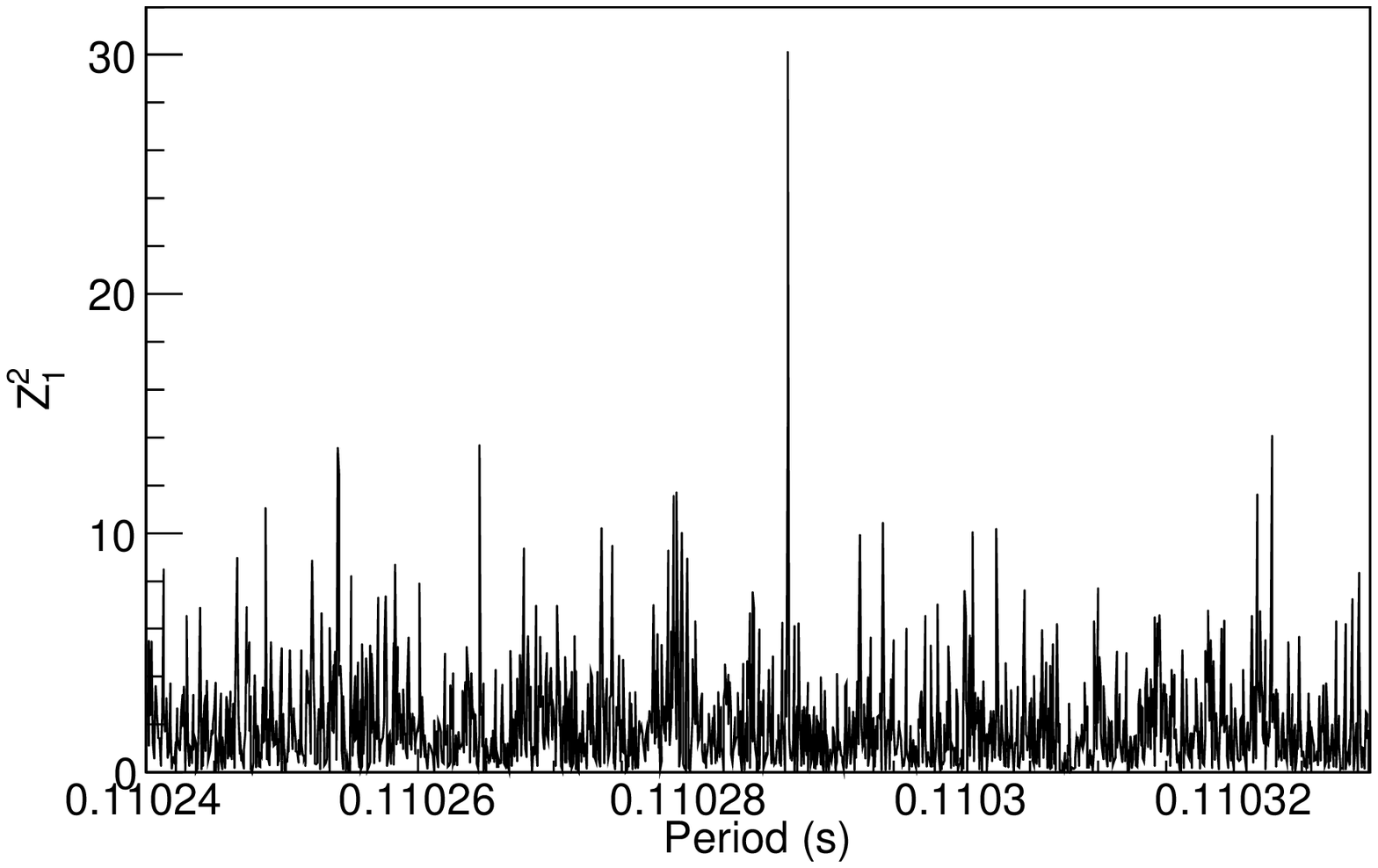}
\includegraphics[scale=0.4]{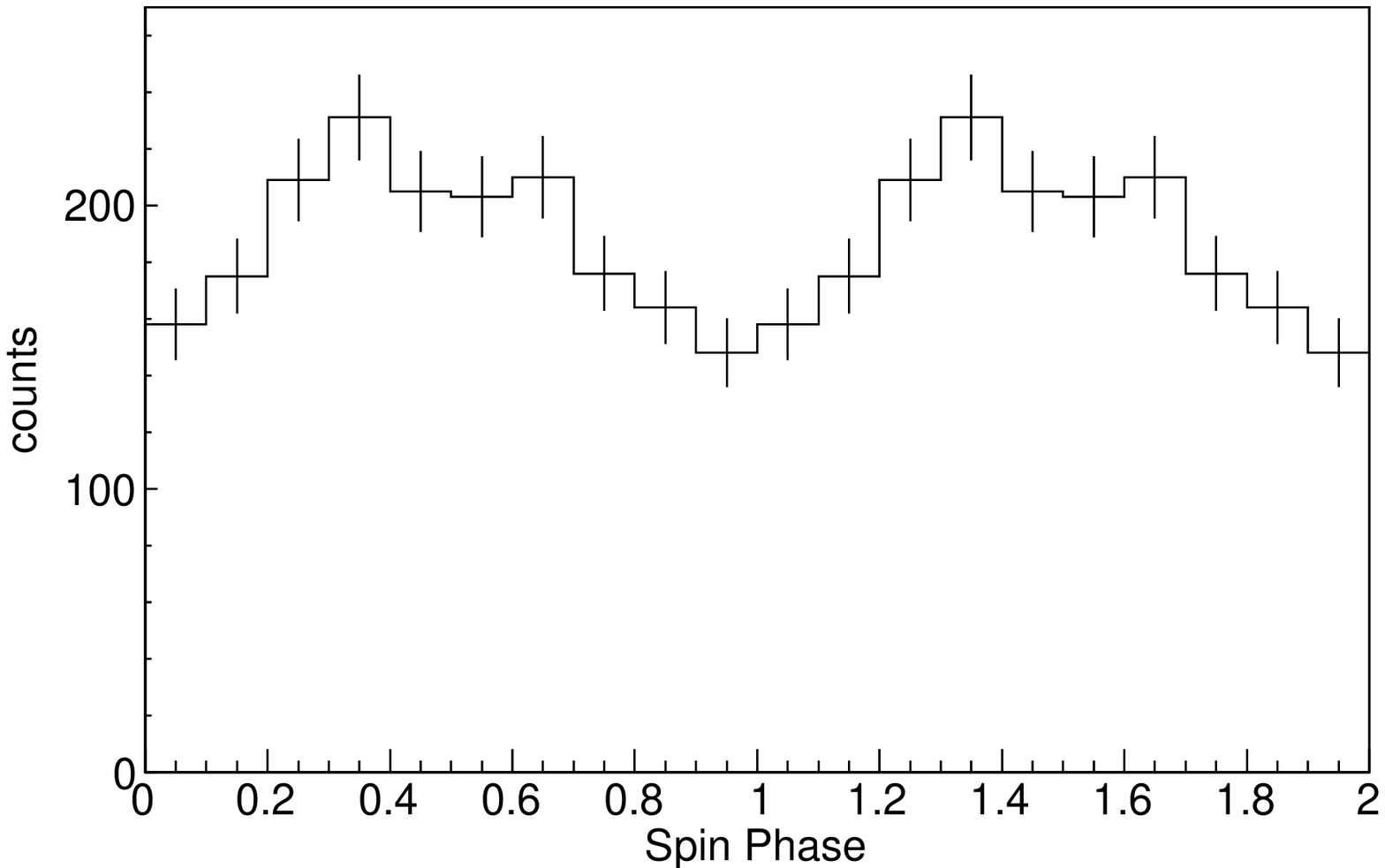}
\caption{{Z$^{2}_{n}$ periodogram (left) and pulse profile folded using the detected period (right). From top to bottom are PSR J1747-2958, PSR J2021+3651 and PSR J1826-1256)}. For the specific data used in each case, see the text.}
\label{Z2}
\end{figure*}
\end{center}
%%%%%%%%%%%%%%%%%%%%%%%%%%%%%%%%%%

%%%%%%%%%%%%%%%%%%%%%%%%%%%%%%%%%%%
\begin{center}
\begin{figure}
\centering
\includegraphics[scale=0.4]{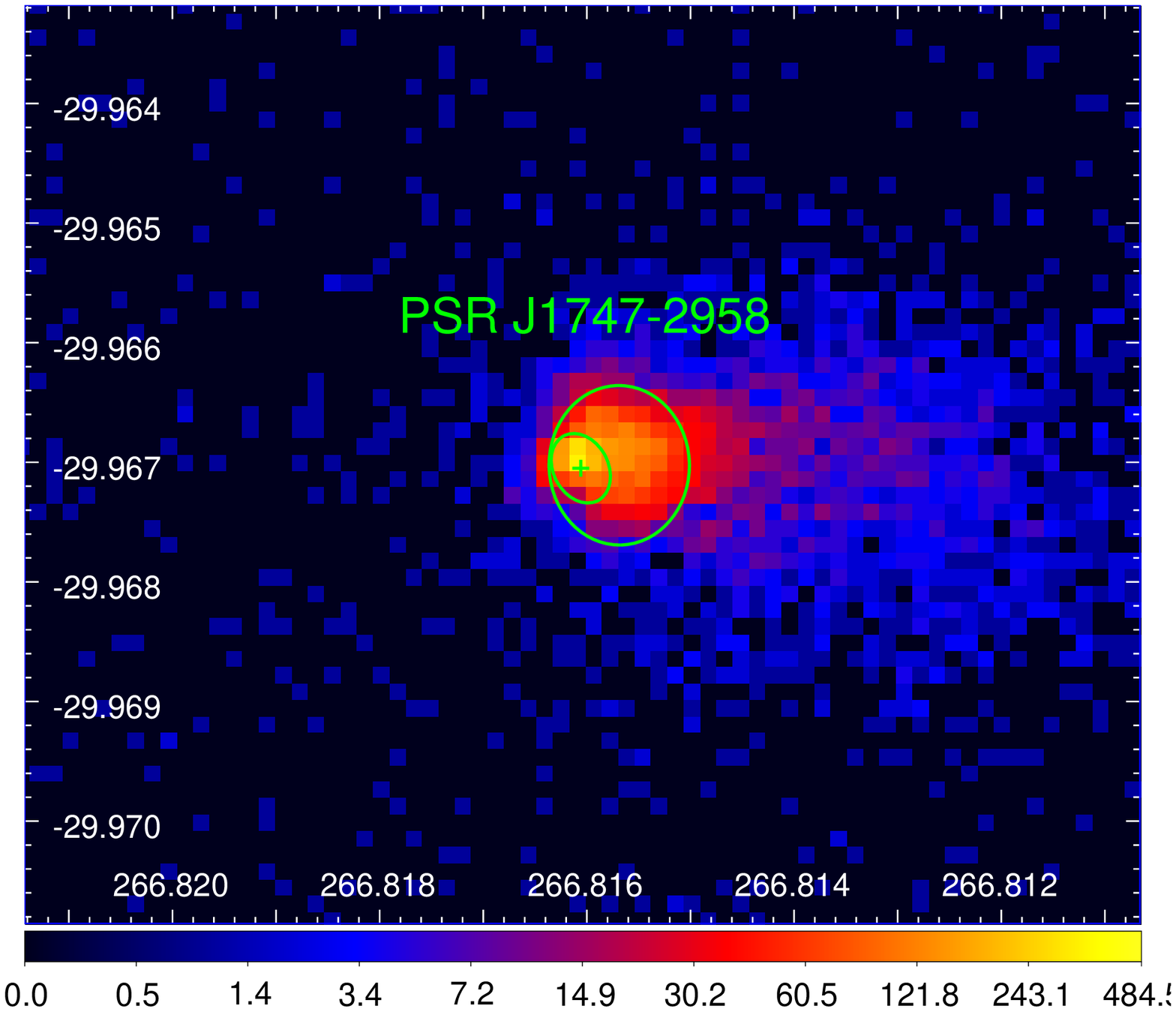}
\includegraphics[scale=0.4]{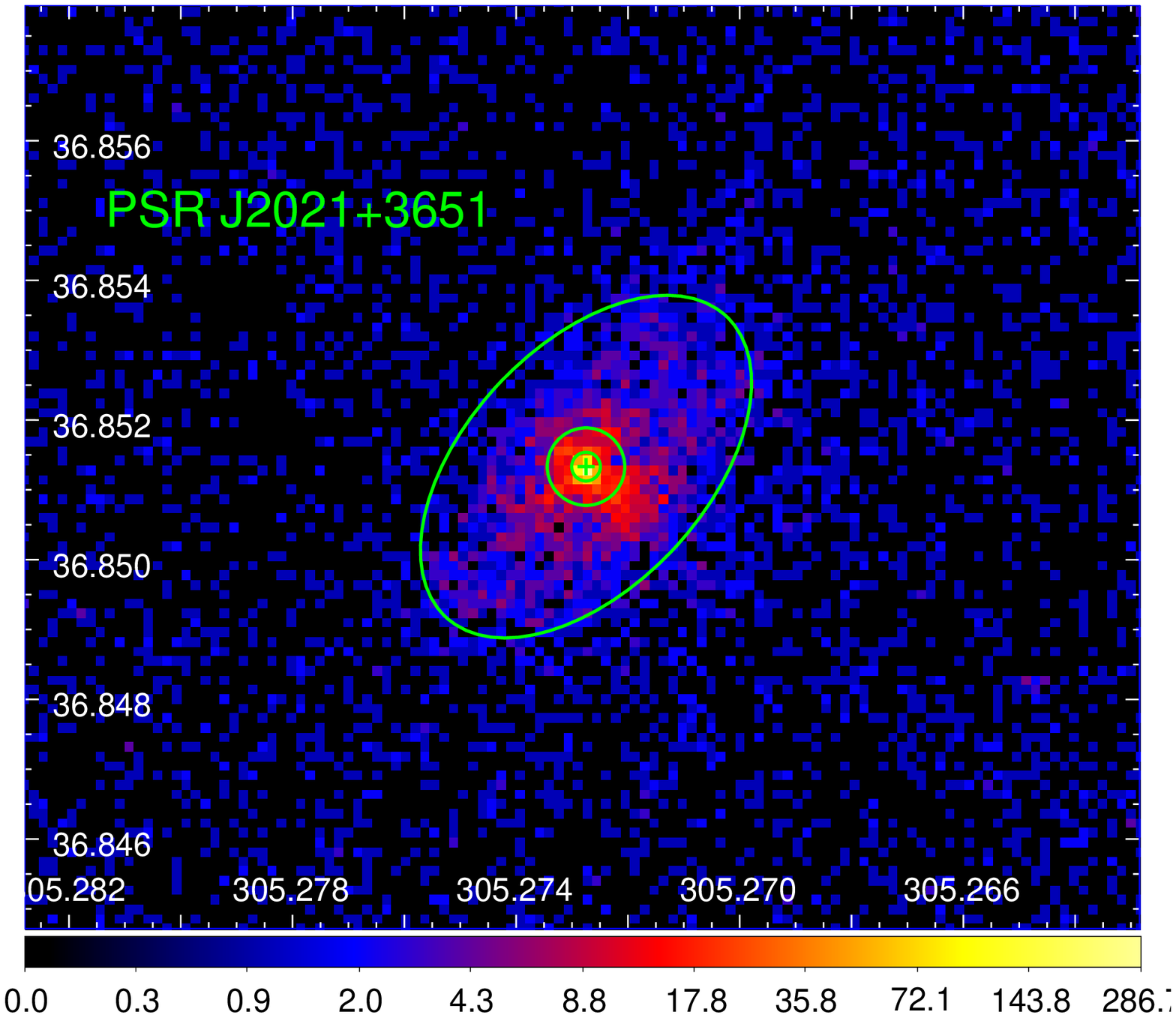}
\includegraphics[scale=0.4]{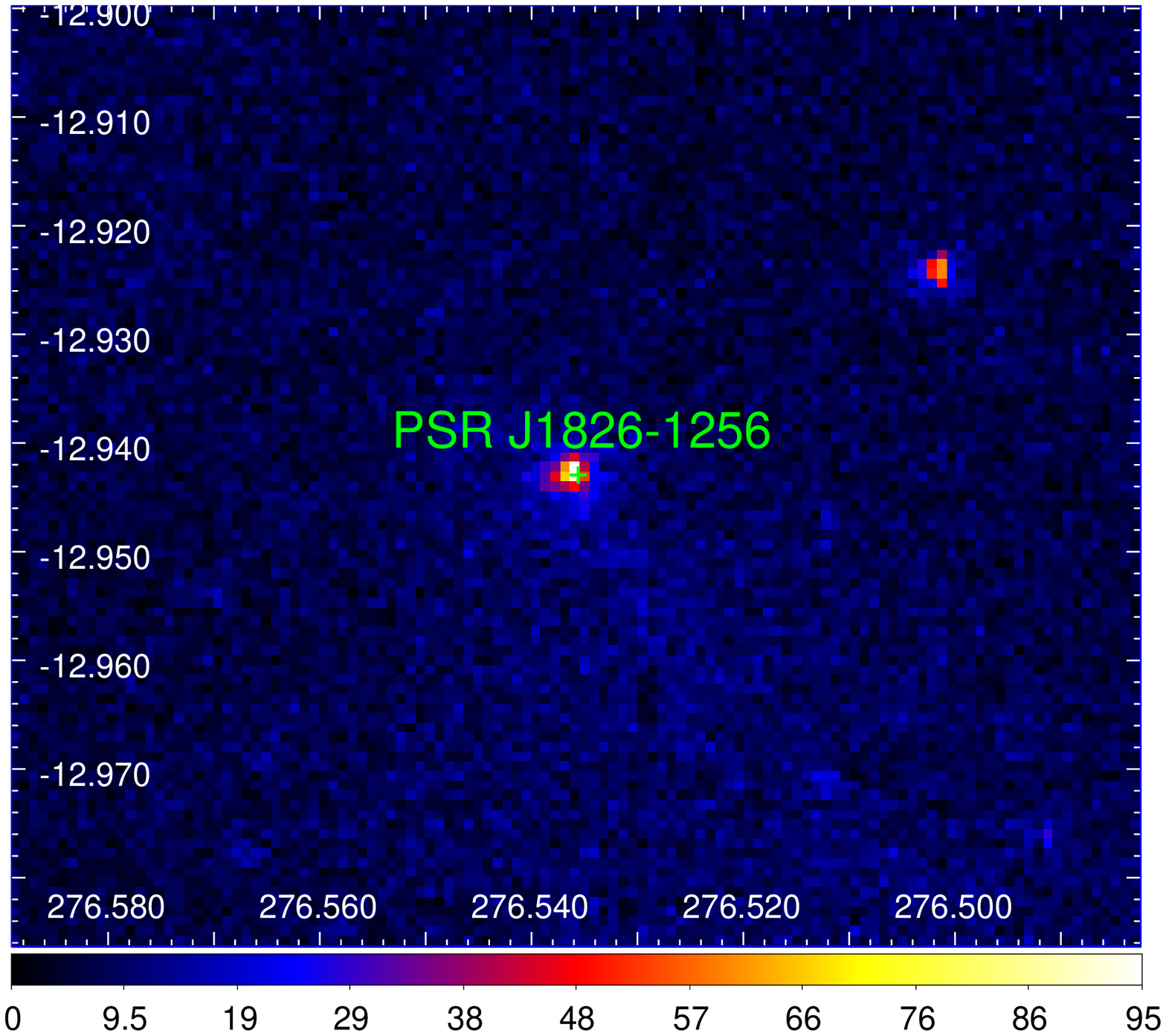}
\caption{Log-scaled counts maps of the region around J1747-2958 (top left, \textit{Chandra}/ACIS-I  image in 0.3-10 keV, obsID 14519), J2021+3651 (top right, \textit{Chandra}/ACIS-S image in 0.3-10 keV, obsID 7603), J1826-1256 (bottom, {\it XMM-Newton}/MOS1 \& MOS2 combined image in 0.2--10 keV, obsID 0744420101).
The positions of the pulsar are shown with green crosses.
The regions {used to extract} the pulsar and nebula contributions in J1747-2958 \& J2021+3651 are shown with ellipses and circles, respectively (see Section 2.1 and 2.2 for details).
The X-and Y-axis are {R.A.} and {decl.} referenced at J2000.}
\label{maps}
\end{figure}
\end{center}
%%%%%%%%%%%%%%%%%%%%%%%%%%%%%%%%%%

%%%%%%%%%%%%%%%%%%%%%%%%%%%%%%%%%%%
\begin{center}
\begin{figure}
\centering
\includegraphics[scale=0.292]{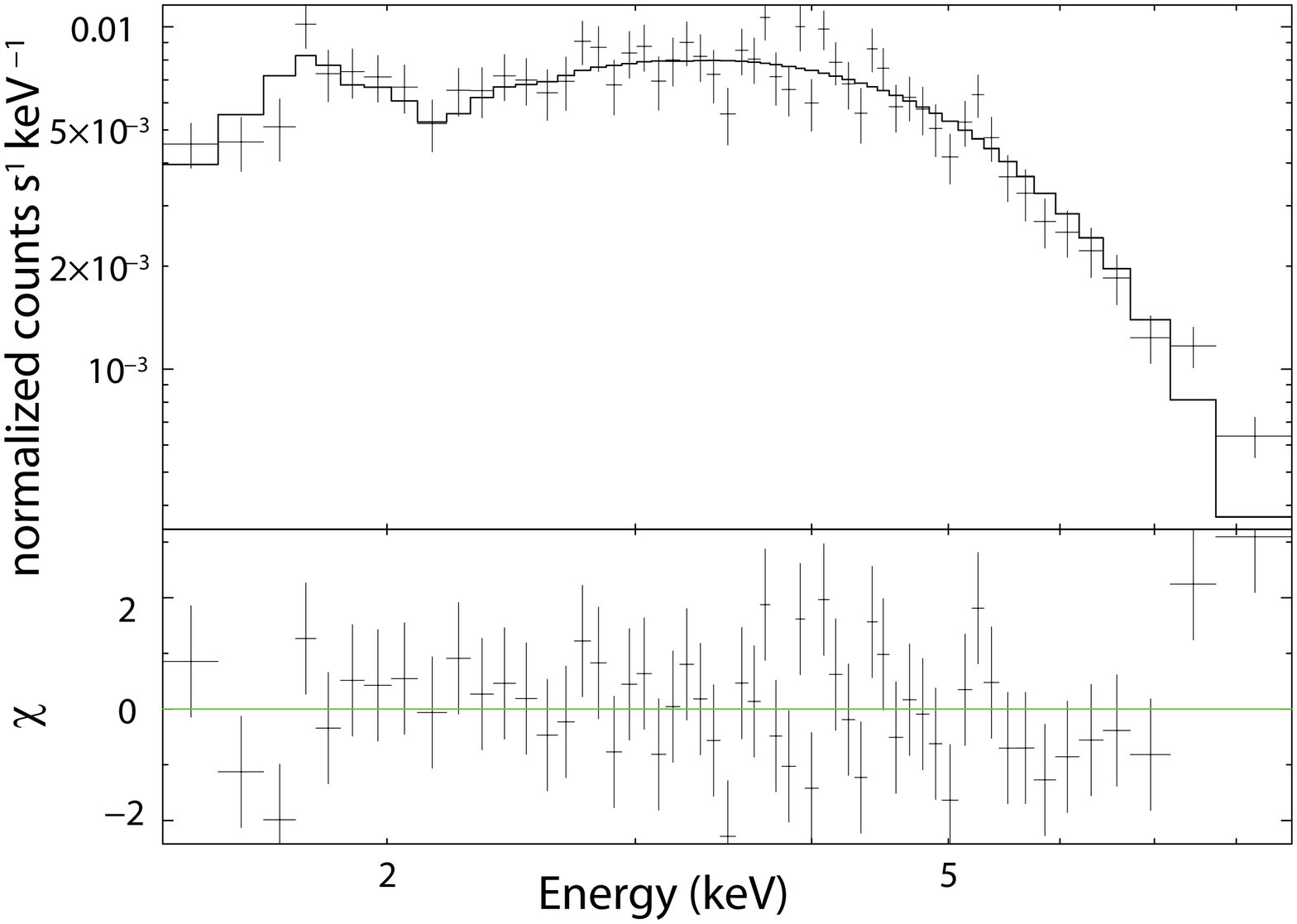}
\includegraphics[scale=0.453]{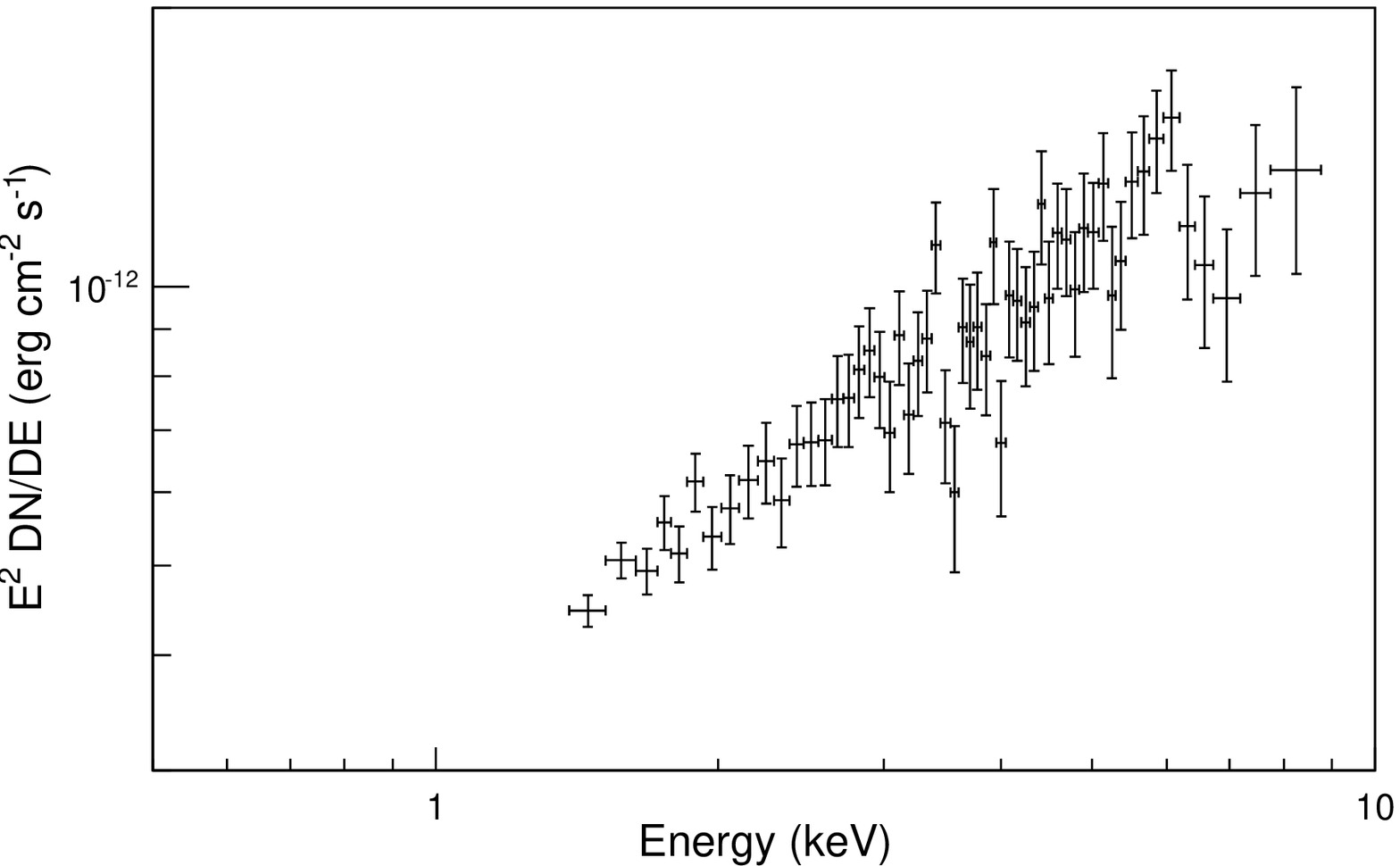}\\
\includegraphics[scale=0.295]{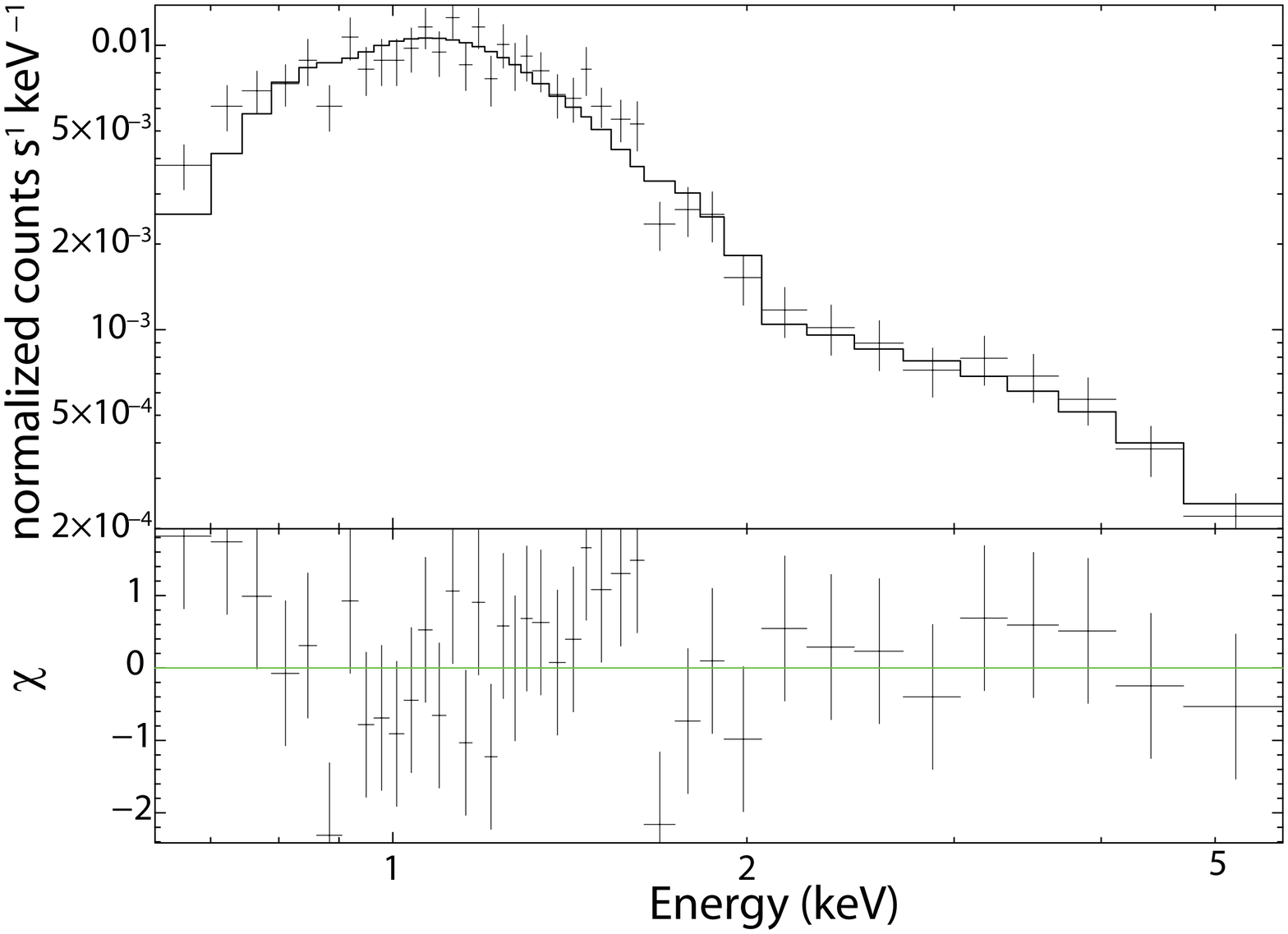}
\includegraphics[scale=0.46]{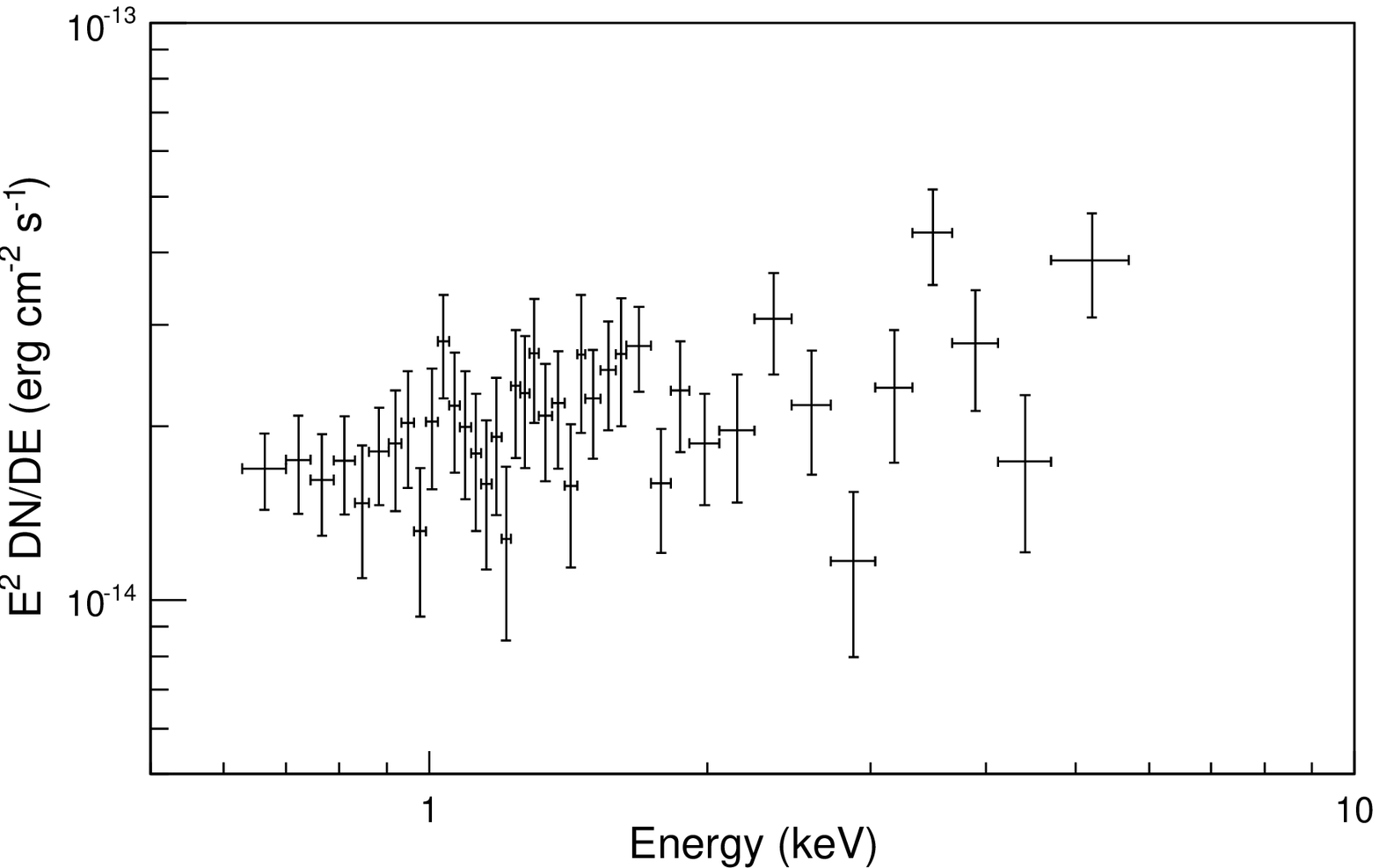}\\
\includegraphics[scale=0.292]{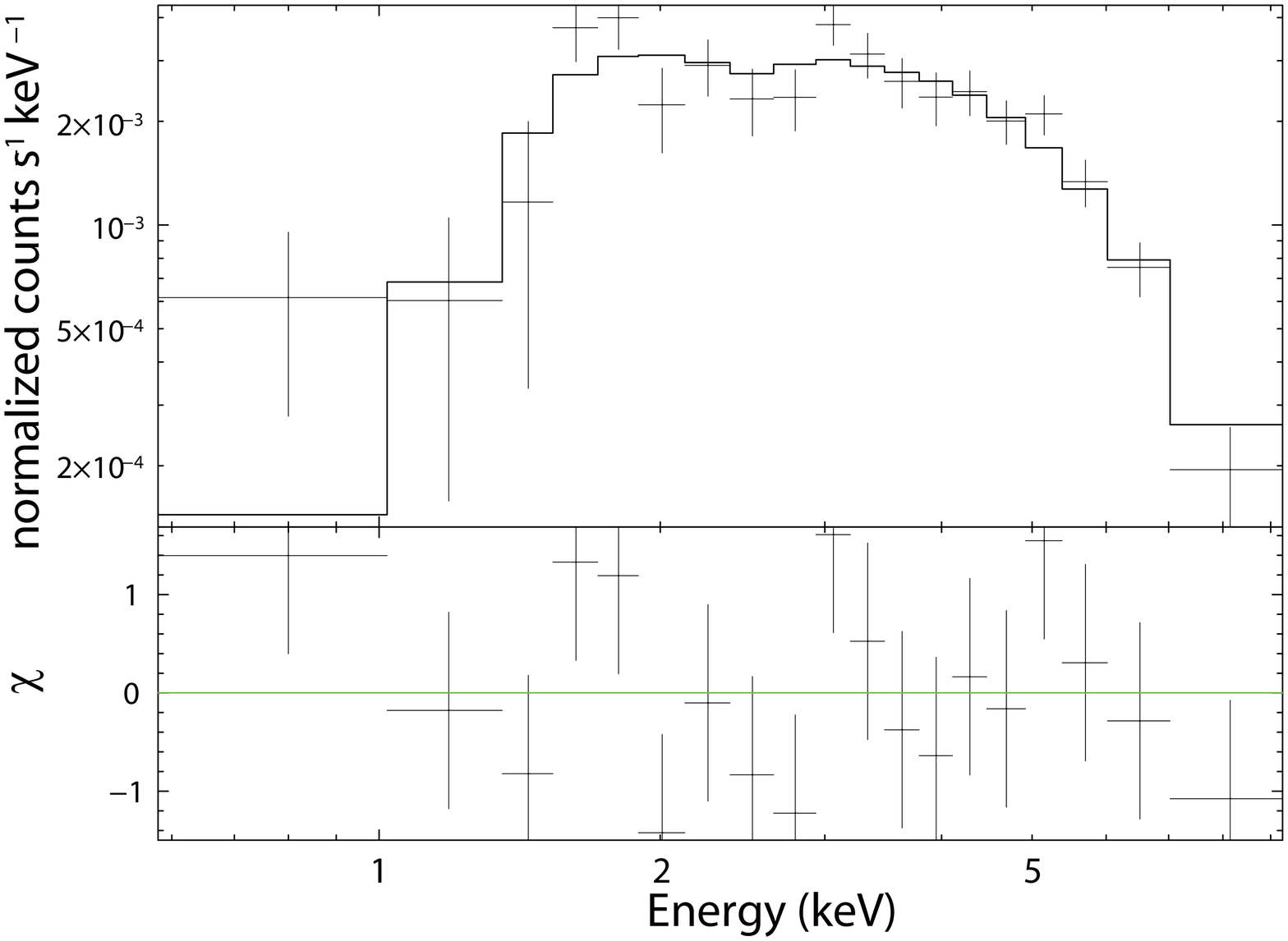}
\includegraphics[scale=0.46]{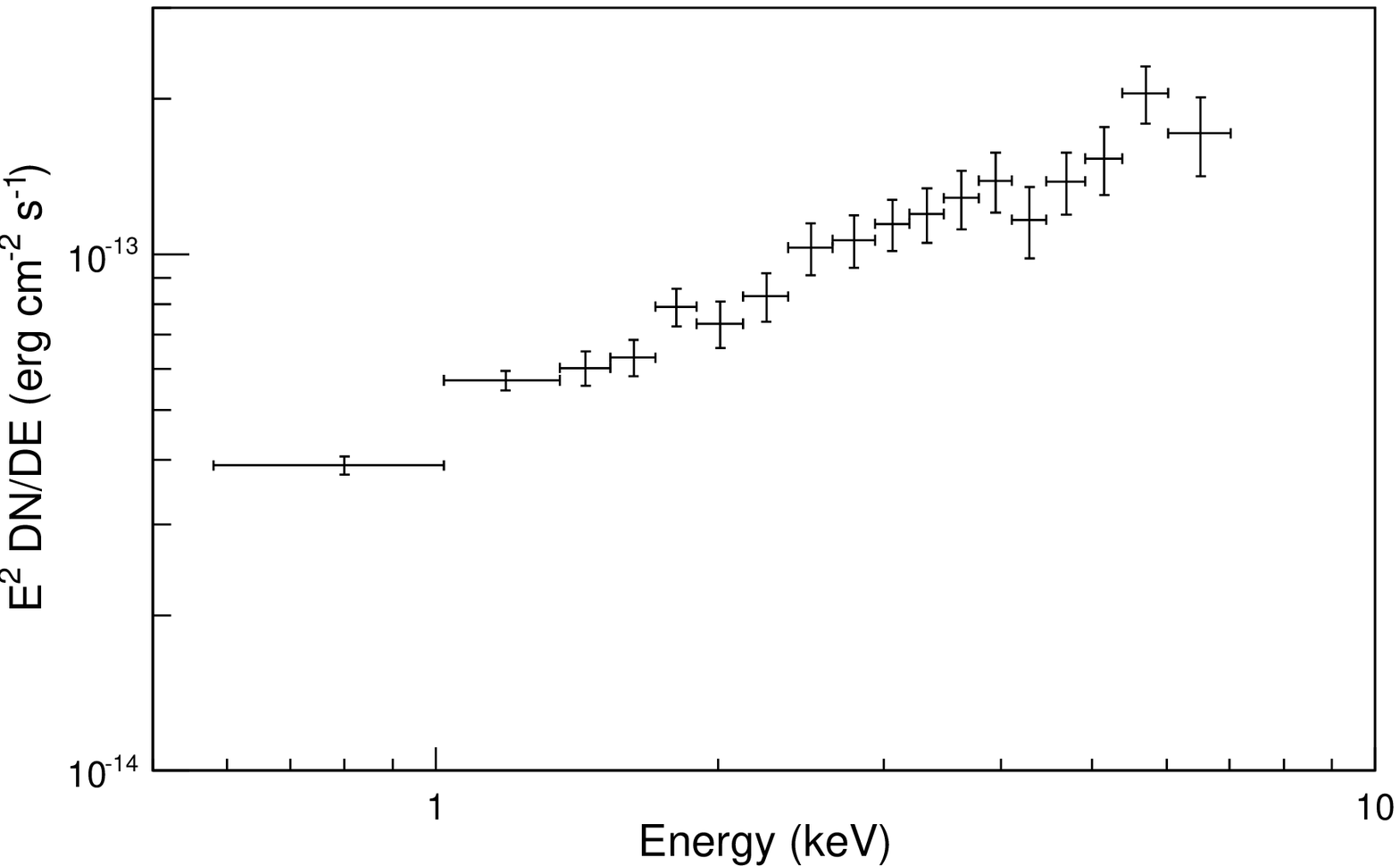}
\caption{
The left panels show the spectrum of J1747-2958 (top),
J2021+3651 (middle), both with {\it Chandra}/ACIS-S data;
and J1826-1256 (bottom), with combined {\it XMM-Newton}/MOS1 \& MOS2.
The best fitted models {and post-fit residuals} are also shown in each case.
The right panels show the corresponding pulsar X-ray spectral energy distributions (only the non-thermal spectral component is shown for J2021+3651).}
\label{spec}
\end{figure}
\end{center}
%%%%%%%%%%%%%%%%%%%%%%%%%%%%%%%%%%

\begin{table*}{}
\centering
\scriptsize
\caption{Spectral fits to the X-ray pulsars}
\begin{tabular}{llllll}
\\
\\
\hline\hline
  Souce name           &    N$_{H}$        &     kT       &Spectral index $\Gamma$  & Unabsorbed flux in 0.2-10 keV      &  $\chi^{2}$/D.O.F. \\
                               &  (10$^{22}$ cm$^{-2}$)    &   (keV)  &                                        & (10$^{-13}$ erg~cm$^{-2}$s$^{-1}$) &    \\
\\
\hline\hline                 % inserts double horizontal lines
\\
J1747-2958         & 2.58$^{+0.25}_{-0.24}$ & - &  1.25 $\pm$0.12              & 23.46 $^{+1.24}_{-1.04}$ & 65.08/56\\
\\
\hline
\\

J2021+3651         & 0.69 (fixed)    & 0.15$\pm$0.01 &  1.72 $\pm$0.30              & 0.78 $^{+0.12}_{-0.09}$ & 38.34/36
\\
%\\
%G75.2+0.1            & 0.69$\pm$0.05    &-&                    1.45$\pm$0.06                  &7.47 $^{-0.17}_{+0.18}$ & 23.46/35
%\\
\\
\hline
\\
J1826-1256        &  2.28$^{+0.49}_{-0.42}$ & - &   1.31$^{+0.22}_{-0.20}$  & 3.32 $^{+0.47}_{-0.31}$  &  17.29/16\\
\\
                          &  0.79$^{+0.31}_{-0.26}$ & 1.54$^{+0.13}_{-0.12}$ &  - & 2.01$\pm$0.12   &  17.46/16
\\
\\
\hline\hline                 % inserts double horizontal lines
\\

\label{psrj_fit}
\end{tabular}
\end{table*}

\subsection{J2021+3651: Confirmed detection of the pulsar in the `Dragonfly' nebula}

PSR J2021+3651 is a 103.7 ms radio and gamma-ray pulsar (Roberts et al. 2002; Abdo et al. 2009) associated with nebula G75.2+0.1 (referred to as the Dragonfly nebula; Hessels et al. 2004; Van Etten et al. 2008).
In this case, hints of X-ray pulsations have earlier been {reported using} {\it Chandra} at a significance of 3.7$\sigma$ (Hessels et al. 2004).
{Here, we re-analyzed this {\it Chandra} observation.}% as well as {\it XMM-Newton} observations.

{{\it Chandra} observed PSR J2021+3651 on Feb. 12th, 2003 with ACIS-S operating in continuous clocking mode (obs. ID 3902).
It provides 20 ks exposure with sufficient timing resolution (2.85 ms).
Adopting the X-ray position from Hessels et al. (2004), we extracted photons with a radius of 5 pixels.
Because of the short exposure and low counts, we analyzed all events in 0.3-10 keV via a similar Z$^{2}_{n}$-test procedure as described before (Buccheri et al. 1983).
A peak at $P=0.10375(7)$ s (90\% uncertainty) is significantly detected with a Z$^{2}_{1}$ value of 26.21, corresponding to a significance $\sim 4.8 \sigma$.
The radio predicted spin period is 0.10372423 s (Hessels et al. 2004), which is consistent with the X-ray detected period.
The folded pulse profile is shown in Fig. \ref{Z2}, leading to a pulse fraction of {19.86\%$\pm$4.27\%.}
The pulse profile is different from that reported in Hessels et al. (2004) which was produced using the radio expected period and X-ray photons in 0.5-3 keV from a extraction radius of 3 pixels.}

We used the archival {\it Chandra} observations (obs. ID 3901, 7603, 8502) for the spectral analysis of PSR J2021+3651,
adopting similar methods to those described in Kirichenko et al. (2015) and Van Etten et al. (2008).
The pulsar spectrum was extracted from all three observations using a source region of radius 1.5 pixels ({0.74 arcsec}).
The nebula spectrum was extracted from an elliptical region with semi-axes of 6.2 and 10.6 arcsec and a position
angle of 137$\degree$.
A circle around the pulsar with a radius of 2 arcsec was excluded from this region (see Figure \ref{maps}, top right panel).
{Corresponding spectra are combined from the different observations to increase statistics, as similarly done for PSR J1747-2958.
}%
{A simultaneous fitting to the spectra from different observations leads to consistent results}.
The spectrum of the nebula was fitted with an absorbed power-law, leading to N$_{H}$={(6.9$\pm$0.5)} $\times$ 10$^{21}$ {cm$^{-2}$}, photon index  $\Gamma$= 1.45$\pm$0.06, and {a} flux level in the 0.2--10 keV band of 7.47$^{+0.17}_{-0.18}$$\times$10$^{-13}$ erg~cm$^{-2}$s$^{-1}$, values which are consistent with previously published results (Kirichenko et al. 2015; Van Etten et al. 2008; Hessels et al. 2004).
As suggested by previous studies, the pulsar was modelled with an absorbed sum of the power-law and blackbody components.
The absorption column density was fixed at the nebula fitted value.
To model the nebula contribution to the pulsar spectra, we added another power-law component to the pulsar spectral model, with a photon index fixed at the nebula value.
Its flux level was fixed at the 5\% of total nebula flux, as suggested by Kirichenko et al. (2015) and Van Etten et al. (2008).
Our fitting results for the pulsar are consistent with those of Kirichenko et al. (2015) and Van Etten et al. (2008) (see Table 1).

\subsection{Detection of PSR J1826-1256}

PSR J1826-1256 is a 110.2 ms radio quiet gamma-ray pulsar discovered by {\it Fermi}-LAT  (Ray et al. 2011).
Its X-ray counterpart has been identified with ASCA and {\it Chandra} (Roberts et al. 2001; Ray et al. 2011) but no X-ray pulsations were reported.

{\it XMM-Newton} observed PSR J1826-1256 with 140 ks exposure, from Oct. 11th to 13th, 2014 (Figure \ref{maps}, bottom panel).
During this observation, PN was operating in small window mode, providing sufficient time resolution {(5.7 ms)} to search for X-ray pulsations.
Using the X-ray position from Ray et al. (2011), we searched for pulsations of J1826-1256 via the Z$^{2}_{n}$-test procedure.
%
%To increase the statistics,
We extract photons from PN data using a radius of 10 arcsec in 0.3--10 keV.
We found a peak at $P=0.11028(5)$ s (90\% uncertainty) (see Figure \ref{Z2}).
The Z$^{2}_{1}$ statistic of this peak is 30.12, which corresponds to a significance {$\sim 5.1 \sigma$.}
%
%{The is no trial in the timing analysis.}
%
We folded the extracted PN data at the detected period and the pulse profile is also shown in Figure \ref{Z2}, yielding a pulse fraction of {18.06\%$\pm$3.26\%.}
The latest {\it Fermi}-LAT gamma-ray ephemeris covers from Aug. 8th, 2008 to Oct. 18th, 2013.
{We extrapolated the gamma-ray ephemeris to the epoch of the X-ray observation.
}%
The extrapolated pulse period is {$P=0.11024444(2)$} s, which is within the uncertainty of the period we detected.
{When extrapolating the gamma-ray ephemeris, timing noise has not been considered.
Based on current LAT Gamma-ray Pulsar Timing Models, the timing noise of PSR J1826-1256
may reach values as large as $\sim$ 4$\times$10$^{-4}$s
during the X-ray observation,
and leads to additional uncertainties in the predicted period.}

PSR J1826-1256 is detected as a point-like source with {\it XMM-Newton}.
The MOS1 \& MOS2 combined image of J1826-1256 in 0.2--10 keV is shown in Figure \ref{maps}.
{The X-ray spectra for J1826-1256 were extracted separately from MOS1 and MOS2 data, and then combined using the task \emph{epicspeccombine} to increase statistics.}
{We also have tested that a simultaneous fitting to the spectra from MOS1 and MOS2 leads to consistent results}.
The background was subtracted, extracted from a source-free region near PSR J1826-1256 (see Figure \ref{maps}).
The pulsar spectrum could be well fitted with an absorbed power-law (see Figure \ref{spec} and Table 1).
An absorbed blackbody could also lead to acceptable fit.
We could not distinguish both models directly with the current statistics.
However, as we note below, the pulsar spectrum is consistent with the non-thermal model prediction.

%%%%%%%%%%%%%%%%%%%%%%%%%%%%%%%%%%%
\begin{center}
\begin{figure*}
\centering
\includegraphics[scale=0.3]{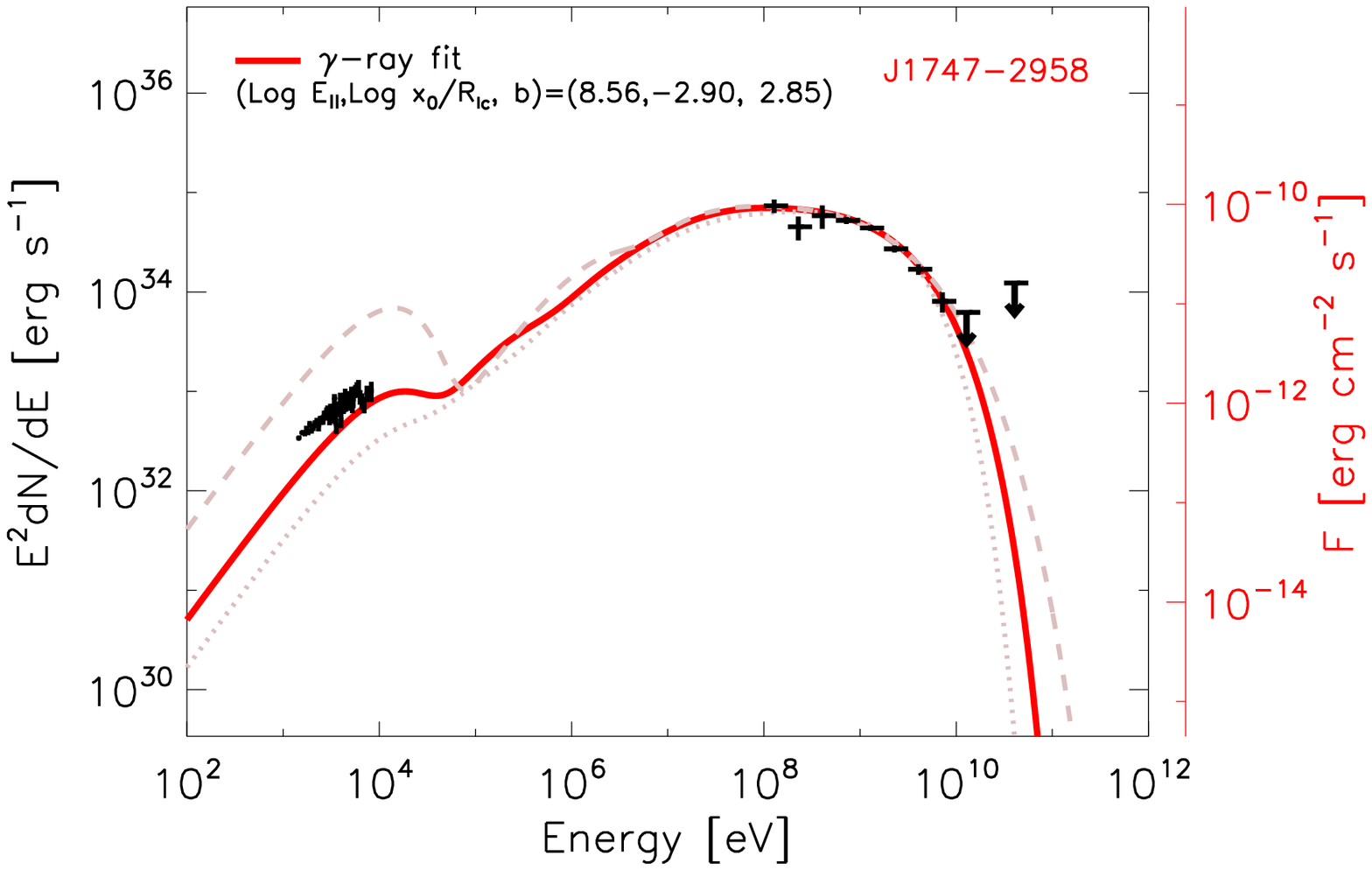}
\includegraphics[scale=0.3]{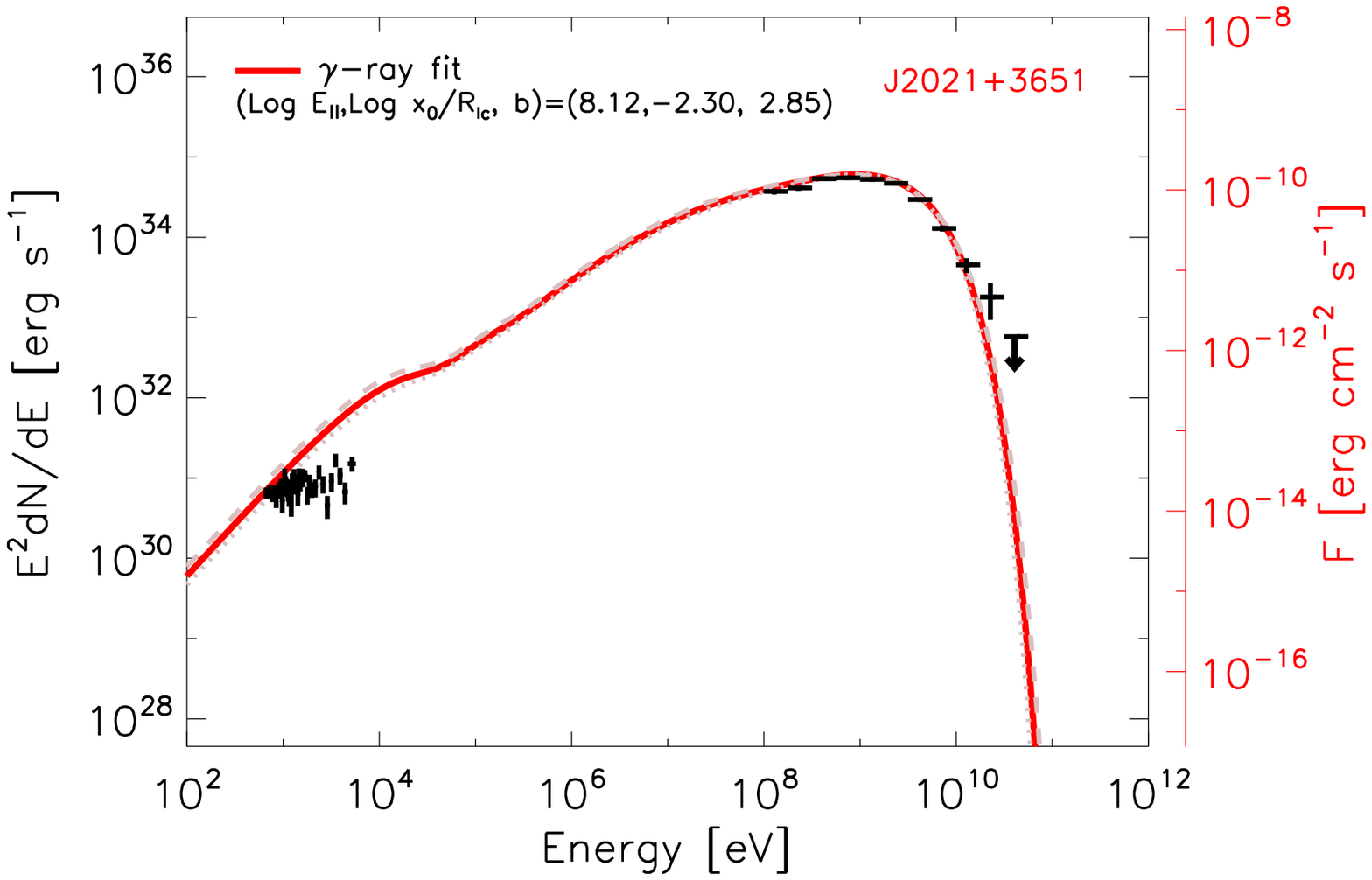}
\includegraphics[scale=0.3]{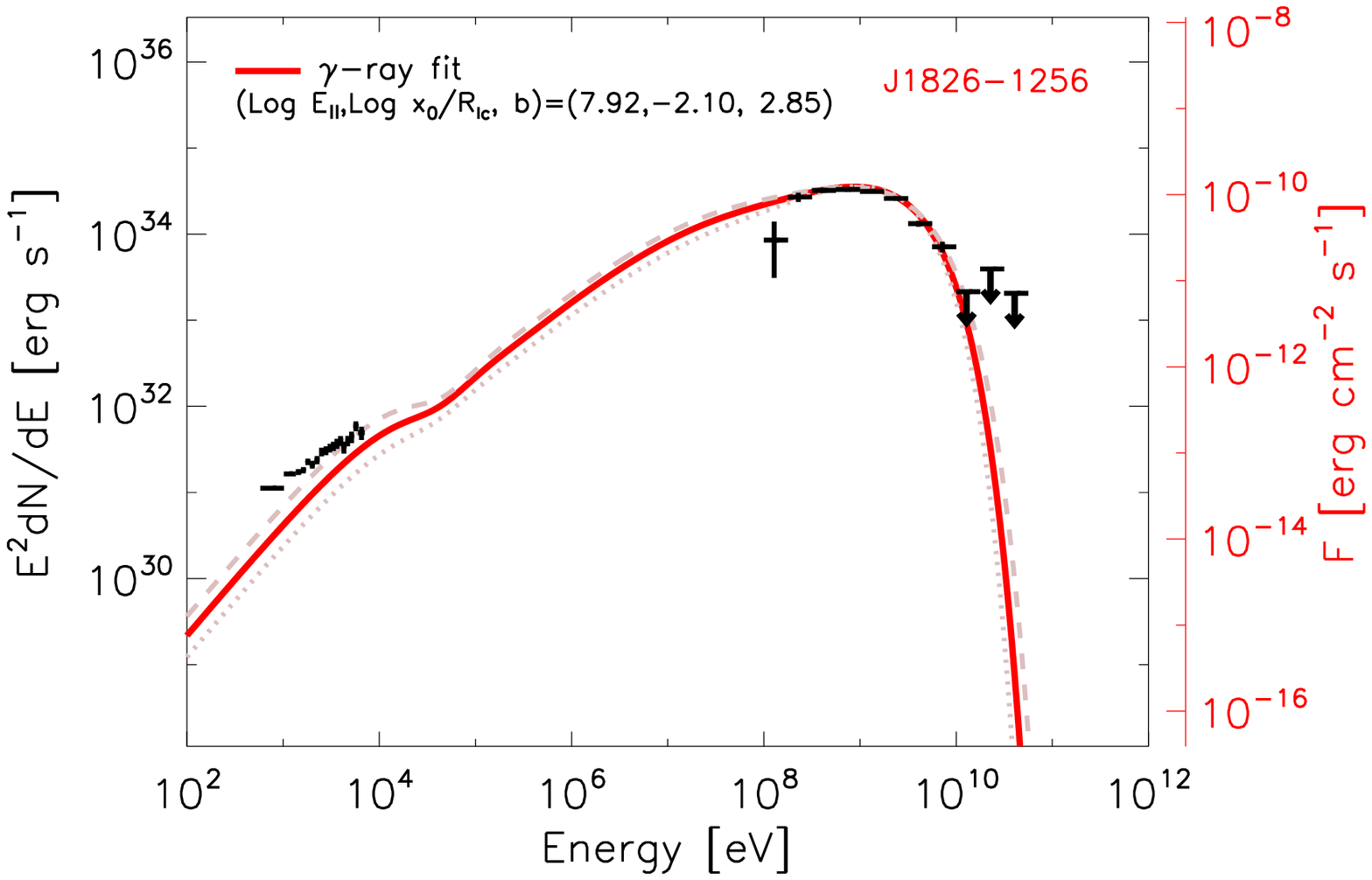}
\includegraphics[scale=0.3]{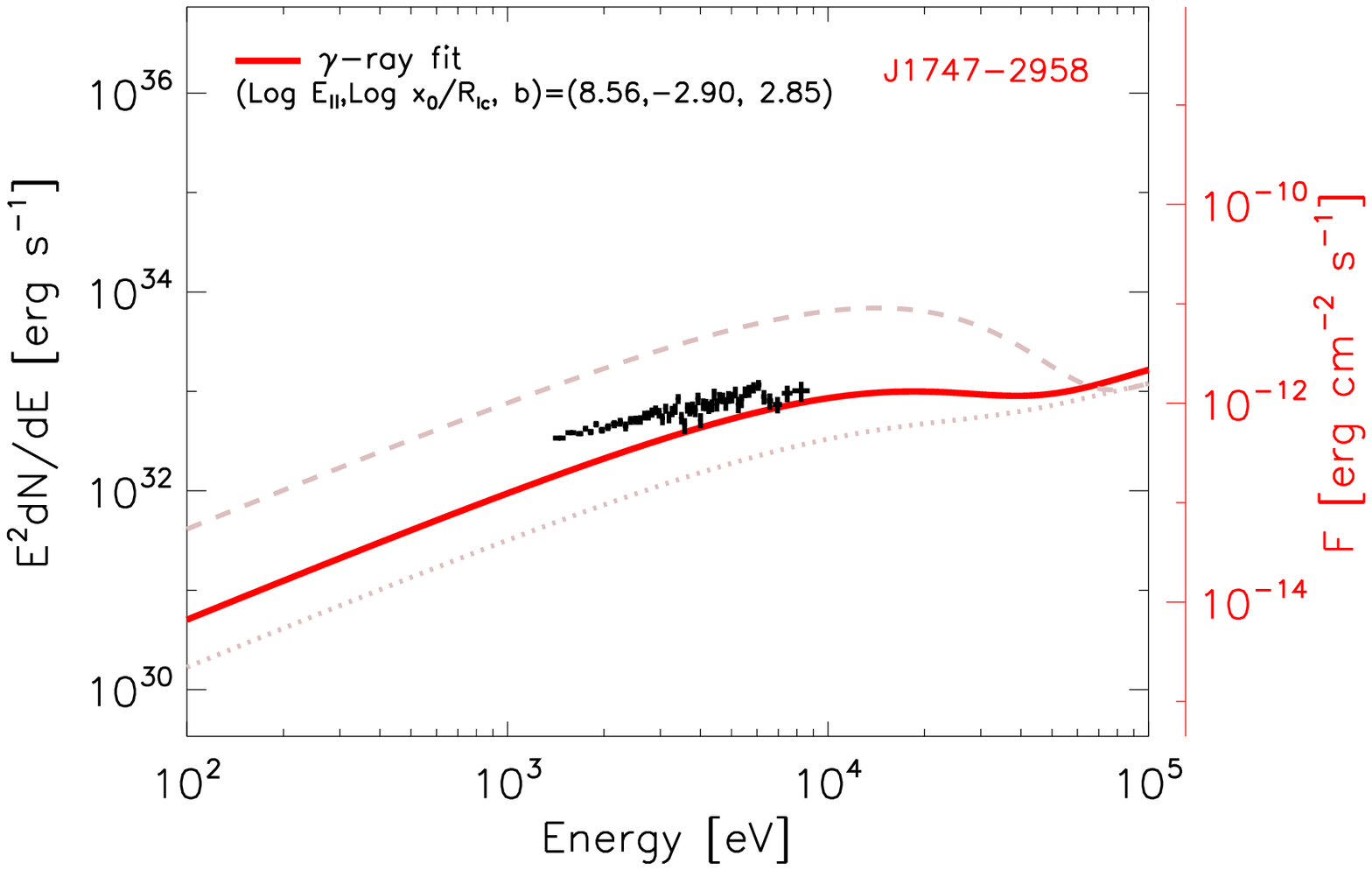}
\includegraphics[scale=0.3]{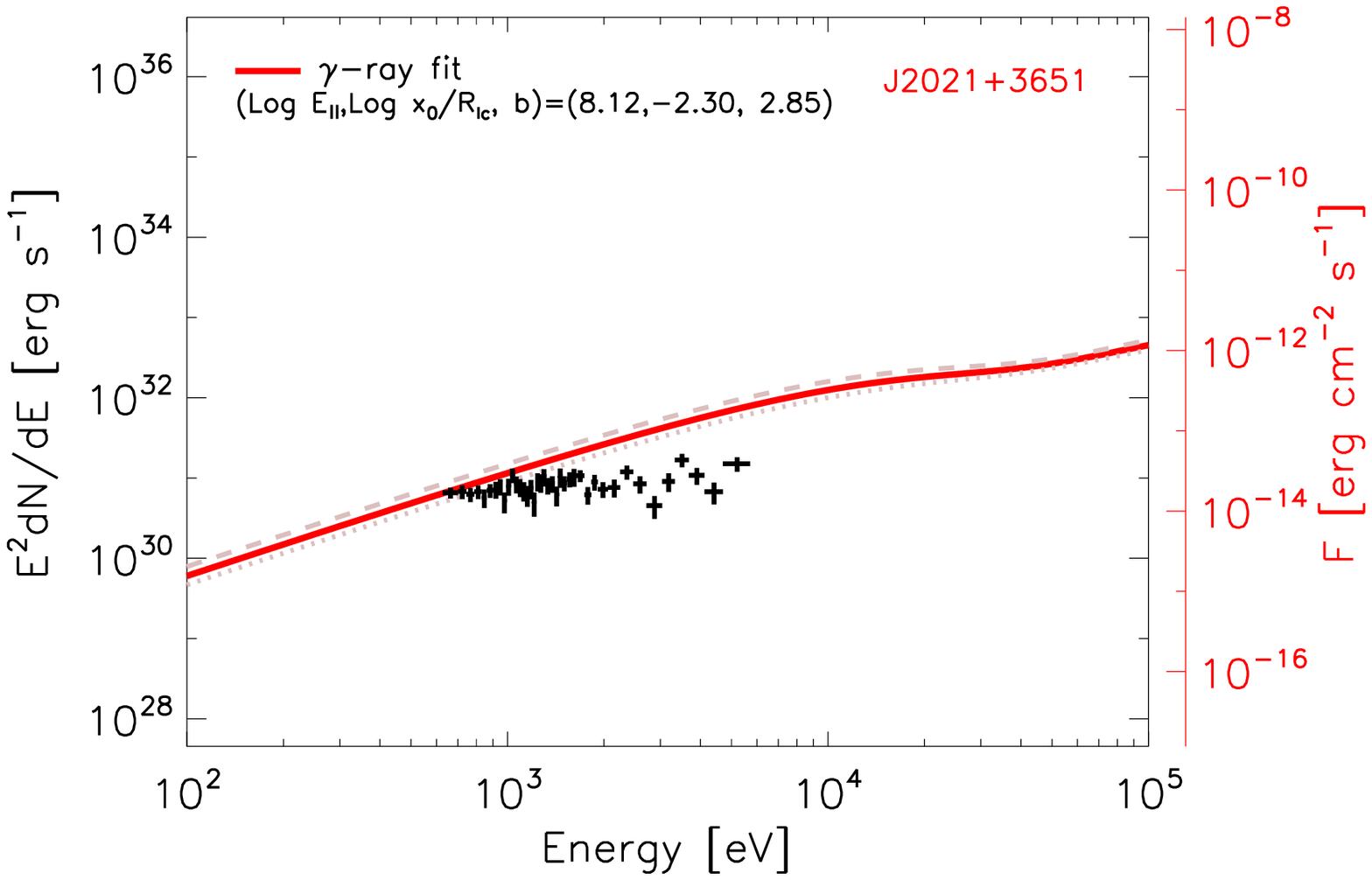}
\includegraphics[scale=0.3]{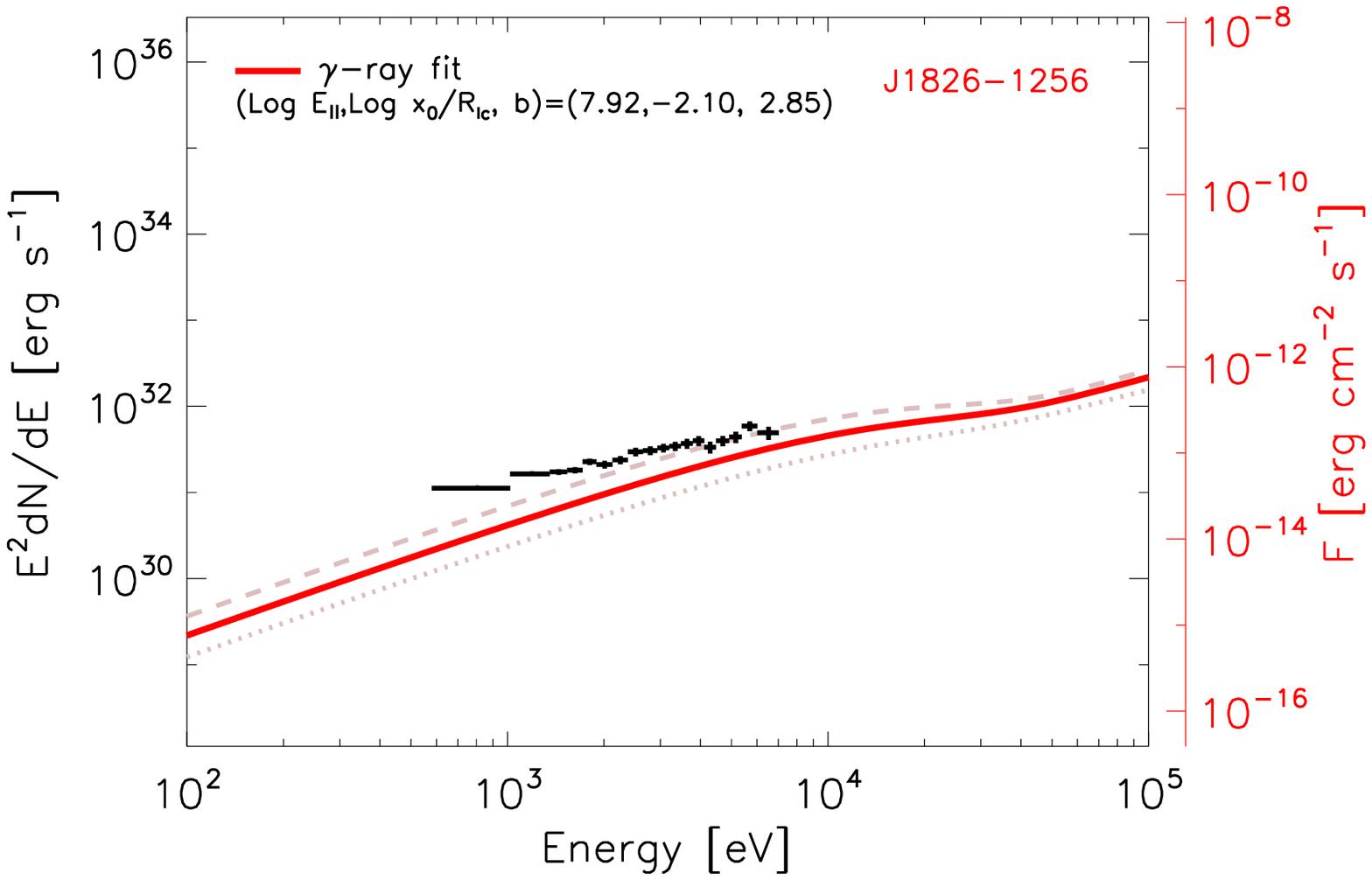}
\includegraphics[scale=0.3]{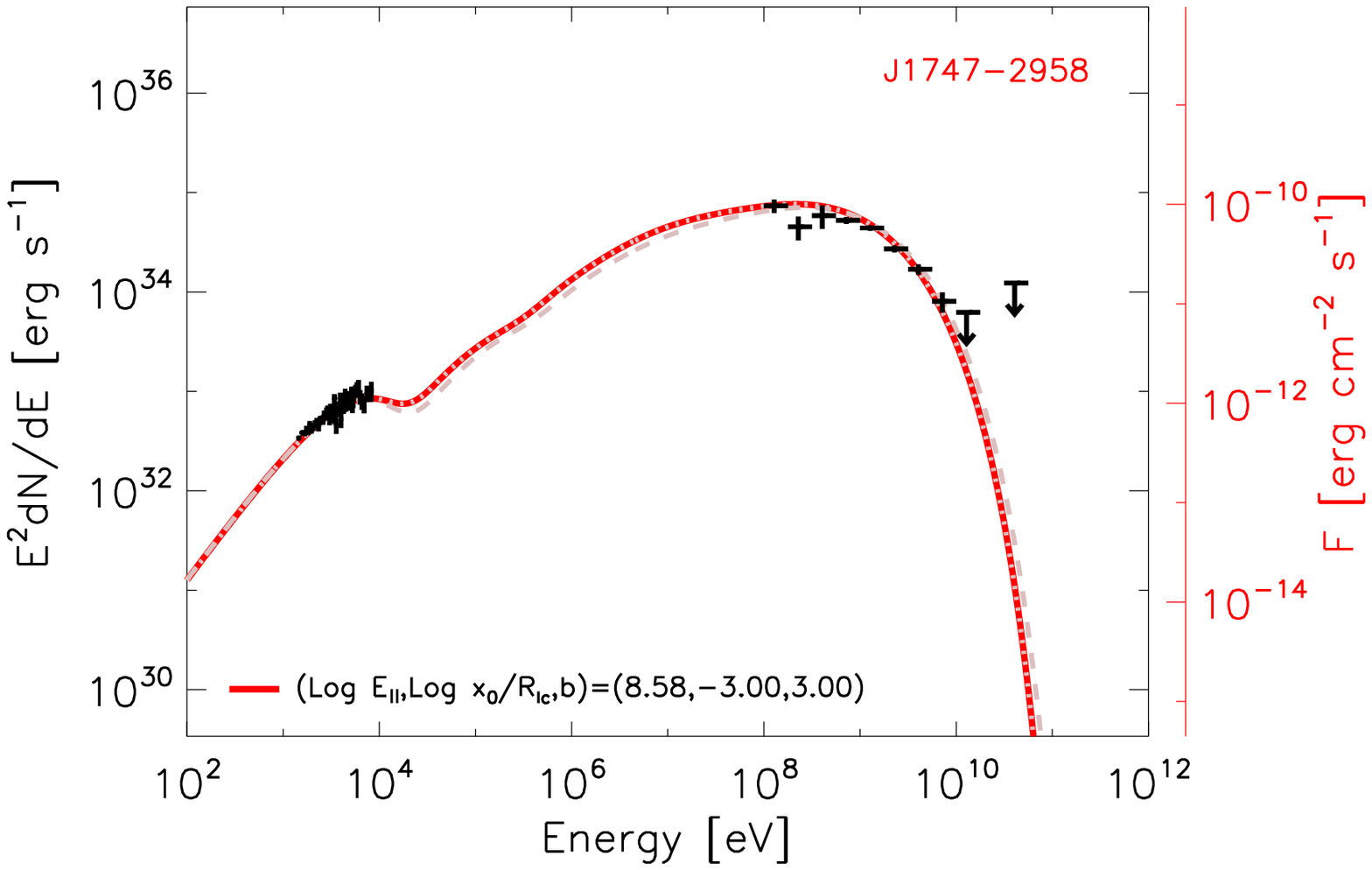}
\includegraphics[scale=0.3]{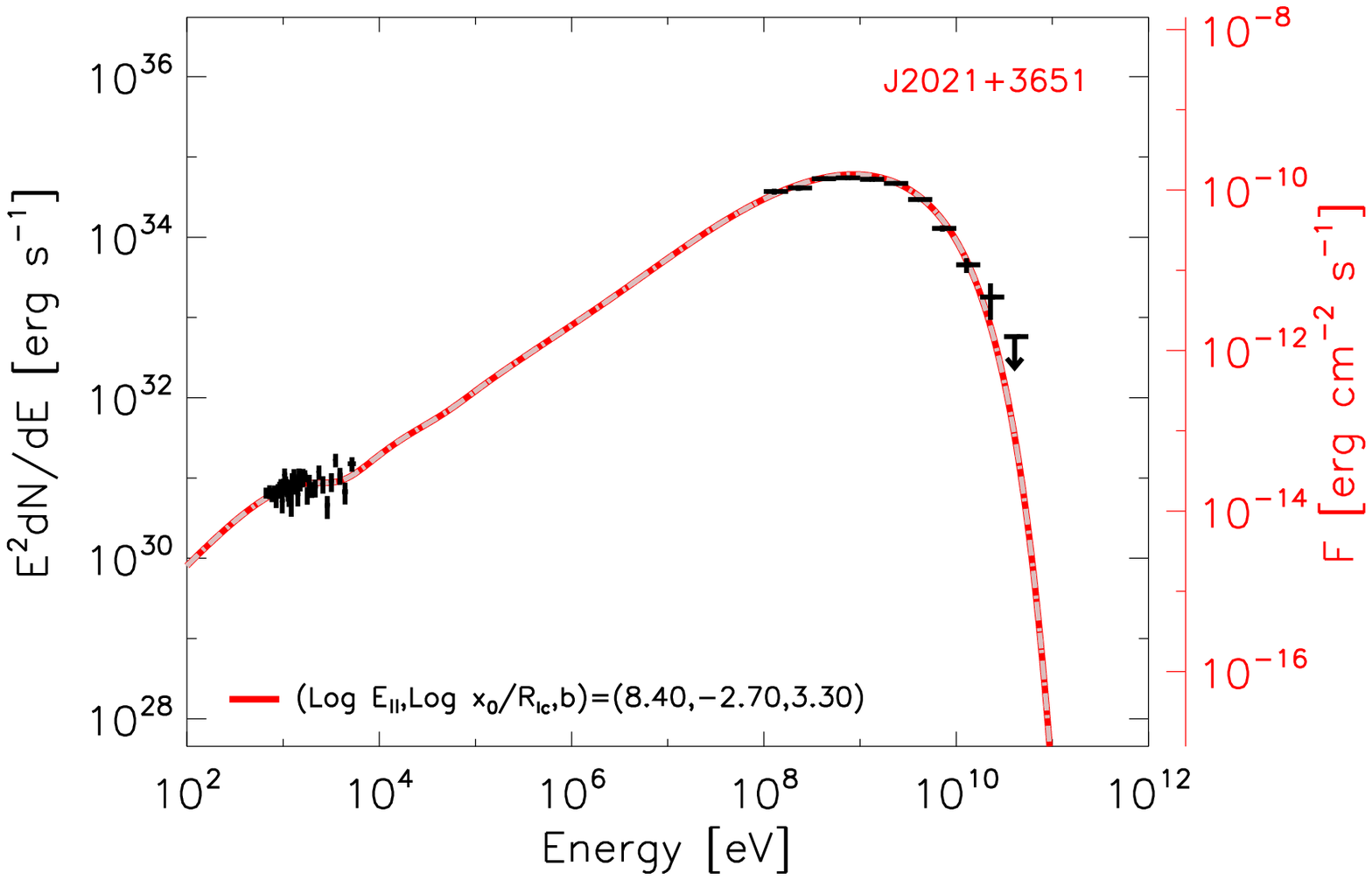}
\includegraphics[scale=0.3]{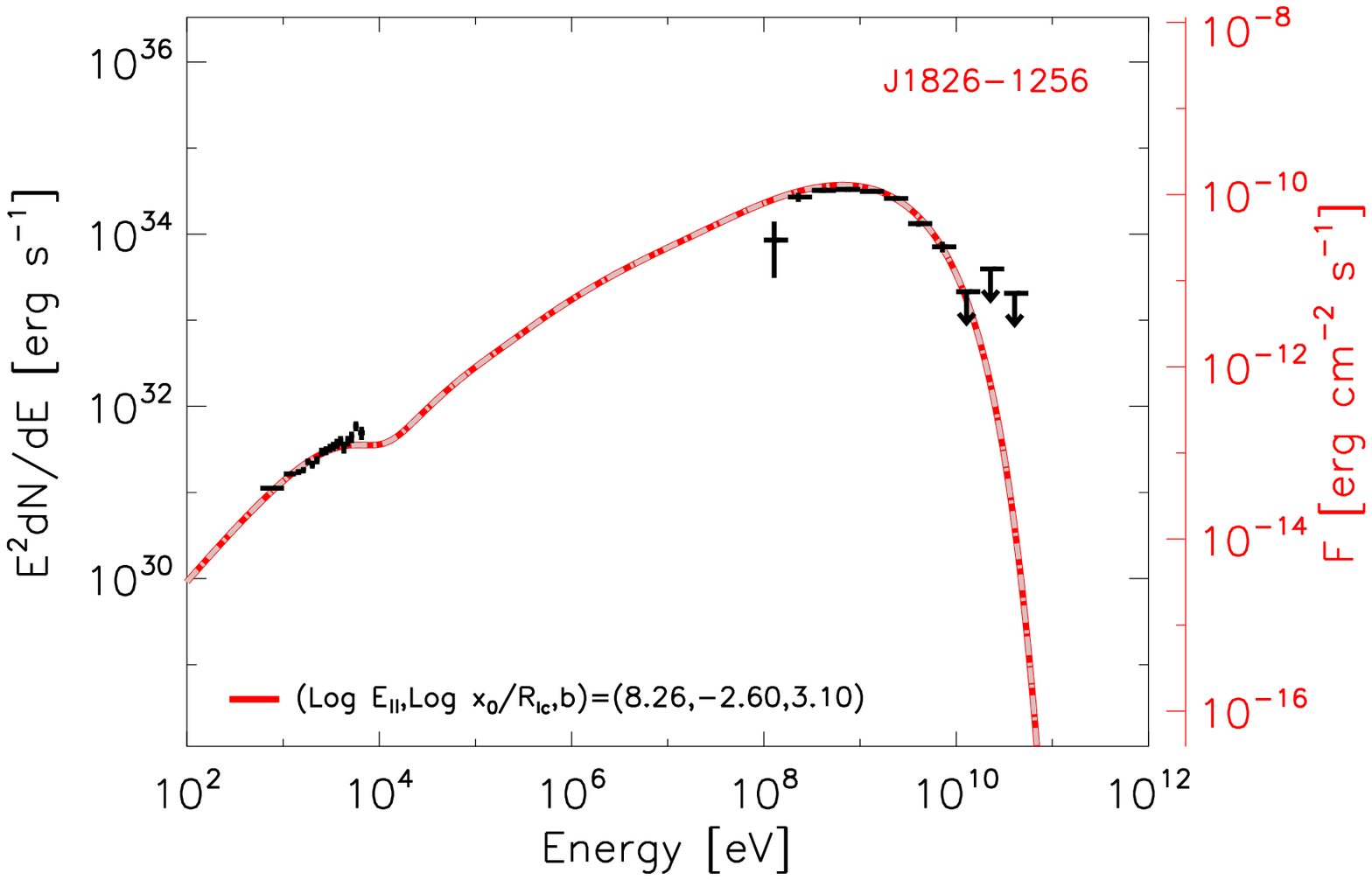}
\includegraphics[scale=0.3]{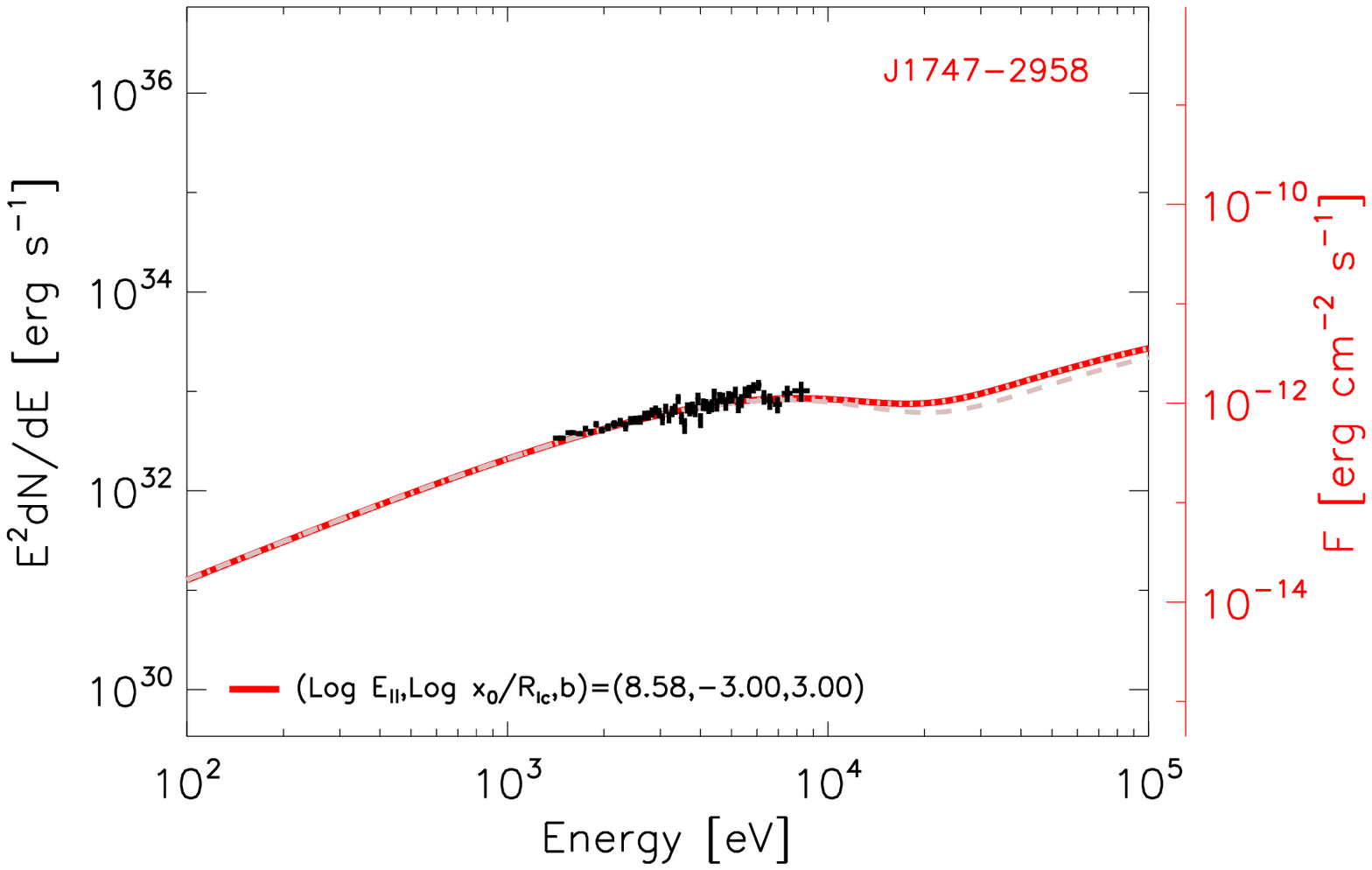}
\includegraphics[scale=0.3]{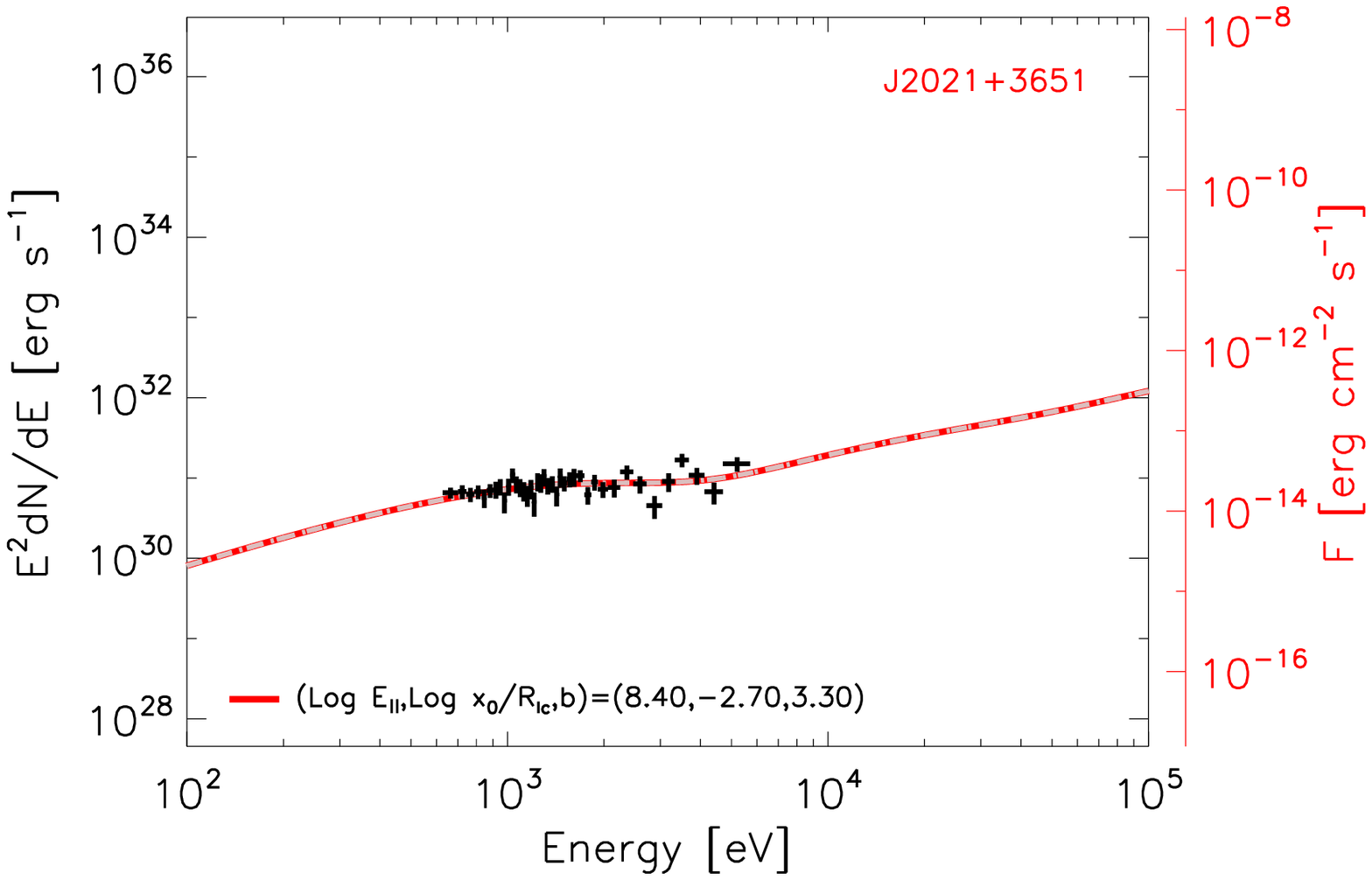}
\includegraphics[scale=0.3]{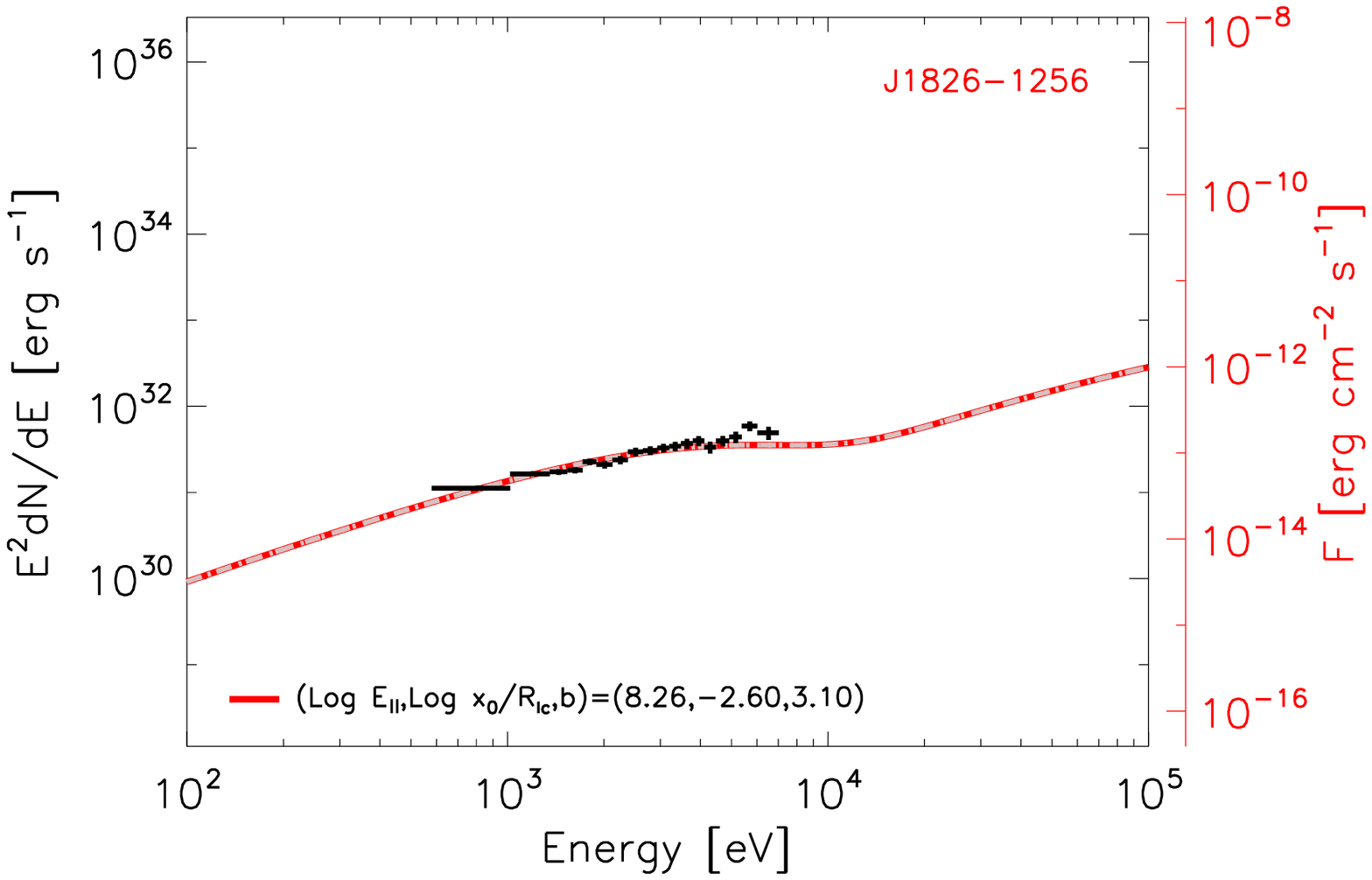}
\caption{The two first rows {show} the unabsorbed non-thermal X-ray spectra as determined in this paper plotted against the model with fixed magnetic gradient ($b=2.85$ see text for an explanation)
used
to select these pulsars as possibly detectable. In the first two rows the X-ray data are not fitted, only the gamma-ray data are.
The second two rows {show} the model fittings considering both the
X-ray and gamma-ray data. The red line is the fitted model in all cases, while the dotted lines (when visible) indicate 1$\sigma$ uncertainty in the fitting parameters.
The second and fourth rows show a zoom of the X-ray band for the corresponding first and third row panels.
The non-thermal X-ray SEDs are taken from the right panel of Fig. \ref{spec}.}
\label{SEDs}
\end{figure*}
\end{center}
%%%%%%%%%%%%%%%%%%%%%%%%%%%%%%%%%%

\section{Discussion}

Fig. \ref{SEDs} shows the derived non-thermal SEDs for all three pulsars in comparison with models.
The first two rows show the initial prediction.
The model is described in detail in Torres (2018) and references therein.
It encompasses two essential ingredients. On one hand, it contains a dynamical, time/position dependent description of particle trajectories in {an accelerating region in the outer part of the magnetosphere}, where particles are subject to radiative losses. On the other hand, it features a computation of the spectrum emitted at each position, while particle traverse the accelerating region.
For both losses and radiation, the full synchro-curvature process is considered (see Vigan\`o et al. 2015a for a description). Just three physical parameters (which we mention below), together with the timing properties of the pulsar (i.e., the measured period and period derivative) define the shape of spectrum.

The red line in each panel of Fig. \ref{SEDs} is the fitted model in all cases.
Panels in the second row show {a zoom of the region in X-ray band} for each of the corresponding first row panels.
For all three panels of the first row
the model is obtained as a fit to the gamma-ray data only.
Such data are lying at energies six orders of magnitude larger than the spectrum we have now determined.
These red lines represent the theoretical prediction we used in order to consider that these three pulsars were actually detectable in X-rays.
In deriving such predictions, the model used only two physical parameters; the accelerating electric field, $E_{||}$, and
the contrast, $(x_0/R_{l})^{-1}$.
The latter is a description of how uniform  is the particle distribution along the accelerating region.
The third physical parameter of the model, the magnetic gradient, $b$, representing a measure of how fast the magnetic field declines along the particle trajectory, was kept fixed.
The value of $b$ in these initial fits was assumed to be 2.85, and not fitted against. This value is the average found for the pulsars detected in non-thermal X-rays (above 20 keV) and gamma-rays studied by Torres (2018). As it was discussed earlier (Vigano et al. 2015b, Torres 2018), having only gamma-ray above 100 MeV makes for a difficult a distinction among different values of $b$.
{The values of these parameters are show in Fig. \ref{SEDs}.
}%
The agreement between the X-ray predictions of these models and the determined spectral data is impressive, {which} confirms that the model works well to select which pulsars among those detected in gamma-rays are detectable in X-rays. This fact can then be used to further enlarge, as we do here, the sample of non-thermal pulsars detected {in the} X-rays, which is still small (see e.g., Kuiper and Hermsen 2015).

The third row of Fig. \ref{SEDs} shows the model fits obtained
considering also the X-ray data, whereas the fourth row zooms into the X-ray region.
These fits have a free value of magnetic gradient, and were obtained spanning uniformly on $E_{||}$, $b$, and $x_0$.
These three physical parameters {provide} a correct description of the whole multiwavelength dataset.
All three fits can cope well with both sets of data in such different energy regimes, confirming that the model is generally applicable.
In all three cases, whereas the values of the accelerating electric field $E_{||}$ and
the contrast $(x_0/R_{l})^{-1}$ are roughly unchanged from the gamma-ray only fits, the chosen magnetic {gradients} are larger.
The value for J2021+3651 ($b=3.30$) is in fact the largest of the magnetic gradients found till now for all X-ray and gamma-ray detected pulsars (see Torres 2018), with all three being comparable to Vela ($b=3.25$), or PSR J2022+3842 ($b=3.10$).
Larger values of $b$ make the spectrum softer at X-ray energies (predicting larger fluxes at soft X-ray energies, see
Supplementary Figure 2 of Torres 2018).
If more normal pulsars would be better described by values of magnetic gradients larger than 2.85, the up-to-now average value, the number of possible X-ray detectable pulsars will increase.
This is something that future studies using new samples of gamma-ray pulsars (e.g. the forthcoming Third {\it Fermi} Pulsar Catalog) should take into account.

\acknowledgments

We acknowledge the support from The National Key Research and Development Program of China (2016YFA0400800), the Spanish grants AYA2015-71042-P, SGR2017-1383, AYA2017-92402-EXP, iLink 2017-1238, and the National Natural Science Foundation of China via NSFC-11473027, NSFC-11503078, NSFC-11673013, NSFC-11733009, NSFC-U1838201, XTP project XDA 04060604 and the Strategic Priority Research Program ¡°The Emergence of Cosmological Structures" of the Chinese Academy of Sciences, grant No. XDB09000000. Jian Li acknowledges the support from the Alexander von Humboldt Foundation. Alessandro Papitto acknowledges the agreements ASI-INAF I/037/12/0 and ASI-INAF 2017-14-H.O, and the Marie Sk\l odowska-Curie grant agreement 660657-TMSP-H2020-MSCA-IF-2014. We acknowledge the support of the PHAROS COST Action (CA16214). We acknowledge discussions with Dr. Long Ji, and help from the XMM-\emph{Newton} \& Chandra Helpdesk staff. Work at NRL is supported by NASA.

\end{document}